\documentclass[aps,prb,twocolumn,superscriptaddress]{revtex4-2}
\usepackage{amsfonts}
\usepackage{graphicx}
\usepackage{epstopdf}
\usepackage{epsfig}
\usepackage{amsmath}
\usepackage[normalem]{ulem}
\usepackage{multirow}
\usepackage{makecell}
\usepackage{array}
\usepackage{booktabs}
\usepackage{mathtools}
\usepackage{bm}
\usepackage{color}
\usepackage{soul,xcolor}
\usepackage{array}
\usepackage{booktabs}

\usepackage{tabularx}
\newcolumntype{Y}{>{\centering\arraybackslash}X}

\usepackage{blindtext}

\begin{document}
\setstcolor{red}

\title{Spin-dependent edge states in two-dimensional Dirac materials with a flat band}

\author{Li-Li Ye}
\affiliation{School of Electrical, Computer and Energy Engineering, Arizona State University, Tempe, Arizona 85287, USA}

\author{Chen-Di Han}
\affiliation{School of Electrical, Computer and Energy Engineering, Arizona State University, Tempe, Arizona 85287, USA}

\author{Ying-Cheng Lai} \email{Ying-Cheng.Lai@asu.edu}
\affiliation{School of Electrical, Computer and Energy Engineering, Arizona State University, Tempe, Arizona 85287, USA}
\affiliation{Department of Physics, Arizona State University, Tempe, Arizona 85287, USA}

\date{\today}

\begin{abstract}

	The phenomenon of spin-dependent quantum scattering in two-dimensional (2D) pseudospin-1/2 Dirac materials leading to a relativistic quantum chimera was recently uncovered. We investigate spin-dependent Dirac electron optics in 2D pseudospin-1 Dirac materials, where the energy-band structure consists of a pair of Dirac cones and a flat band. In particular, with a suitable combination of external electric fields and a magnetic exchange field, electrons with a specific spin orientation (e.g., spin-down) can be trapped in a class of long-lived edge modes, generating resonant scattering. The spin-dependent edge states are a unique feature of flat-band Dirac materials and have no classical correspondence. However, electrons with the opposite spin (i.e., spin up) undergo conventional quantum scattering with a classical correspondence, which can be understood in the framework of Dirac electron optics. A consequence is that the spin-down electrons produce a large scattering probability with broad scattering angle distribution in both near- and far-field regions, while the spin-up electrons display the opposite behavior. Such characteristically different behaviors of the electrons with opposite spins lead to spin polarization that can be as high as nearly 100\%.

\end{abstract}

\maketitle

\section{Introduction}
Dirac electron optics can be demonstrated by the behaviors of ballistic
electrons in the paradigmatic graphene p-n junction system~\cite{betancur:2019}.
Specifically, due to the relativistic quantum phenomenon of Klein tunneling and
the gapless Dirac cone dispersion relation, the transmission of Dirac electrons
through the p-n junction interface resembles a highly transparent focusing lens
with negative refractive index~\cite{xu:2018}, corresponding to a Vaselago
lens~\cite{cheianov:2007} for chiral Dirac fermions in graphene. It provides
an experimental approach to tuning the refractive index through varying the
gate potential, making it possible to realize graphene-based electronic
lens~\cite{cserti:2007} and graphene transistors~\cite{wang:2019}. In Dirac
electron optics, various electronic counterparts of optical phenomena have been
achieved such as Fabry-P\'erot resonances~\cite{shytov:2008,rickhaus:2013},
cloaking~\cite{gu:2011}, Dirac fermion microscope~\cite{boggild:2017},
electron Mie scattering~\cite{heinisch:2013,caridad:2016,gutierrez:2016,lee:2016,sadrara:2019}. In addition, in the framework of Dirac electron optics, diverse
unconventional relativistic quantum phenomena such as anti-super-Klein
tunneling in phosphorene p-n junctions~\cite{betancur:2019} and tilted energy
dispersion effect~\cite{nguyen:2018} have been studied. A rigorous
semiclassical theory beyond the standard WKB approximation for the
two-dimensional (2D) Dirac equation was developed~\cite{reijnders:2018} as the
foundation of Dirac electron optics. Experimentally, Dirac fermion flows were
imaged through a circular Veselago lens using the polarized tip of a scanning
gate microscope~\cite{brun:2019} and nanoscale quantum electron optics was
tested in graphene with atomically sharp p-n junctions~\cite{bai:2018}.

The magnetic exchange field(MEF) provides a natural testbed for the spin-dependent Dirac electron optics. It can be induced by the adjacent magnetic insulator, i.e. 2D-material/magnetic insulator, incorporating EuS~\cite{wei:2016}, and ferromagnetic insulator(FMI)~\cite{singh:2017}, and it enables the efficient control~\cite{haugen:2008,yang:2013} of spin generation and modulation in 2D-materials. Moreover, the MEF in magnetic multilayers is promising to achieve tens or even hundreds of tesla~\cite{li:2013,wei:2016}. The spin-dependent electronic spin lens~\cite{moghaddam:2010}, i.e. the counterpart of the photonic chiral metamaterials, generated by the spin-resolved negative refraction Klein tunneling, has been discussed with the magnetic exchange field in graphene normal-ferromagnetic-normal configuration. It spurs the growth of research about electron optics~\cite{grivet:2013,batson:2002,chen:2016,tian:2012,xu:2018,wang:2019chaos,schrepfer:2021,wang:2022}. In Dirac quantum chimera state~\cite{xu:2018} with MEF interaction, the unusual spin-resolved coexistence states by classically chaotic and integral optical quasibound states have been discovered in the annular cavity made with pseudospin-1/2 Dirac fermions, which have the features about the enhancement of Dirac electron spin polarization.

In this work, we explore spin-dependent edge modes in pseudospin-1 Dirac materials by the electrostatic field and MEF interaction. In the previous work~\cite{xu:2019,XHL:2021}, a class of robust edge modes arises that can resist even fully developed classical chaos and Klein tunneling~\cite{xu:2019,XHL:2021} - a unique feature of pseudospin-1 Dirac materials in the absence of a magnetic exchange potential so that the real spin degrees of freedom are degenerate. (It is plausible that such edge modes possess certain topological features~\cite{XL:2020a,XL:2020b}.) Based on this, we demonstrate that systems of pseudospin-1 Dirac materials with a flat band represent an intriguing manifestation of the coexistence in electronic quasibound states of classic lensing(integrable or chaotic states) and non-classical edge states. The former displays electron-optic scattering and the latter demonstrates unconventional scattering. Between that, the interplay features have been explored, and it achieves nearly 100\% spin polarization.

\begin{figure} [ht!]
\centering
\includegraphics[width=0.8\linewidth]{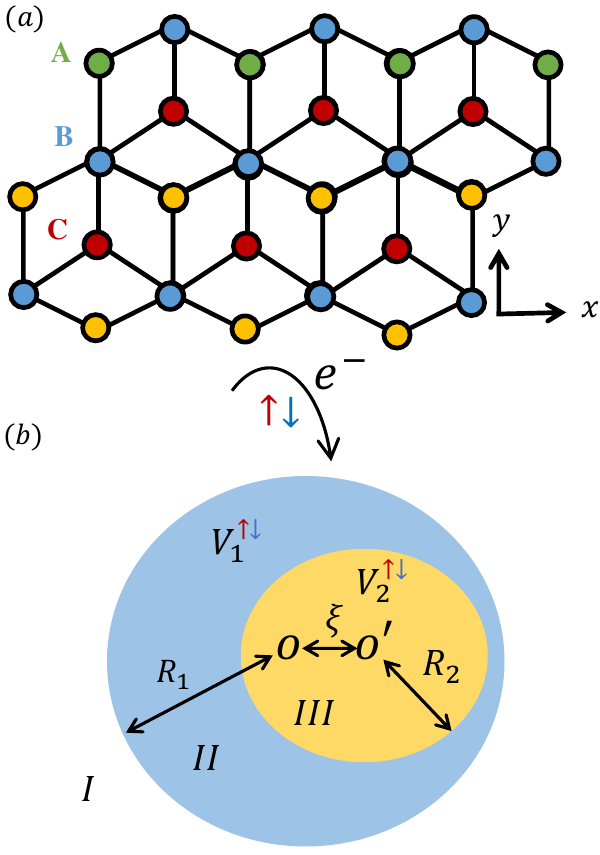}
\caption{Configuration of eccentric circular cavity made by a 2D pseudospin-1 lattice. (a) Dice lattice, one of the possible materials to realize the pseudospin-1 Dirac Weyl Hamiltonian, whose energy band structure consists of a pair of 
Dirac cones and a flat band and (b) a concrete device configuration, and spin-dependent potentials 
$V^{\uparrow\downarrow}_i$ ($i=1,2$) generated by the gate voltages $\nu_1$ and
$\nu_2$ and the
magnetic exchange potential $\mu_1=\mu_2=\mu$, which are applied to the blue and yellow domains, respectively. The radii of the
two circular regions are $R_1=r_0$ and $R_2=0.6 r_0$ with $r_0\sim 100 $ nm being the 
characteristic length.}
\label{fig:config}
\end{figure}

Compared with graphene, pseudospin-1 Dirac material systems are capable of delivering unconventional physical phenomena such as super-Klein 
tunneling~\cite{fang:2016}, novel conical 
diffraction~\cite{mukherjee:2015,vicencio:2015,diebel:2016} and chaos Q-spoiling defiance with edge states~\cite{xu:2019}. An example of pseudospin-1 materials is the dice 
lattice, as shown in Fig.~\ref{fig:config}(a), where the quasiparticles
can be described by the generalized 2D Dirac-Weyl Hamiltonian~\cite{malcolm:2016}. 
Consider an eccentric circular cavity of dice lattice consisting of a large circle 
and a small circular domain inside the large one, where the centers of the two 
circles do not coincide, as illustrated in Fig.~\ref{fig:config}(b). The real 
spin degree of freedom of electron carries becomes relevant when the whole device is placed on a ferromagnetic  
substrate~\cite{wei:2016,singh:2017}, as described by a magnetic 
exchange potential in the Hamiltonian. Now apply two distinct gate 
voltages to the cavity: one to the large circular domain excluding the small 
circle and another to the small circular domain. With appropriate combinations
of the magnetic exchange field strength and the gate voltages, the quantum scattering 
behaviors of spin-up and spin-down electrons can be characteristically 
distinct. For example, spin-up electrons can exhibit lensing modes while  
spin-down electrons would focus on the edge of the large cavity. As a result, the 
spin-down electrons produce a large scattering probability with broad scattering 
angle distribution in both the near-field and far-field regions, while the spin-up 
electrons display the opposite behavior. Such characteristically different 
behaviors of the electrons with opposite spins lead to spin polarization that 
can be as nearly 100 \% We note that the edge modes for spin-down electrons 
break the ray-wave correspondence and confine the electrons for a relatively 
long time~\cite{xu:2019}. In contrast, the lensing modes for spin-up electrons 
have a classical correspondence in the small wavelength limit and they tend to 
leak from the cavity in a short time. In the discussion section, we also explore the spin-resolved quantum scattering of the chaotic scattering and edge mode scattering in the classically chaotic stadium cavity.

Our main code is uploaded to GitHub: https://github.com/liliyequantum/Spin-dependent-edge-states-in-two-dimensional-Dirac-materials-with-a-flat-band.

\begin{figure*} [ht!]
\centering
\includegraphics[width=0.9\linewidth]{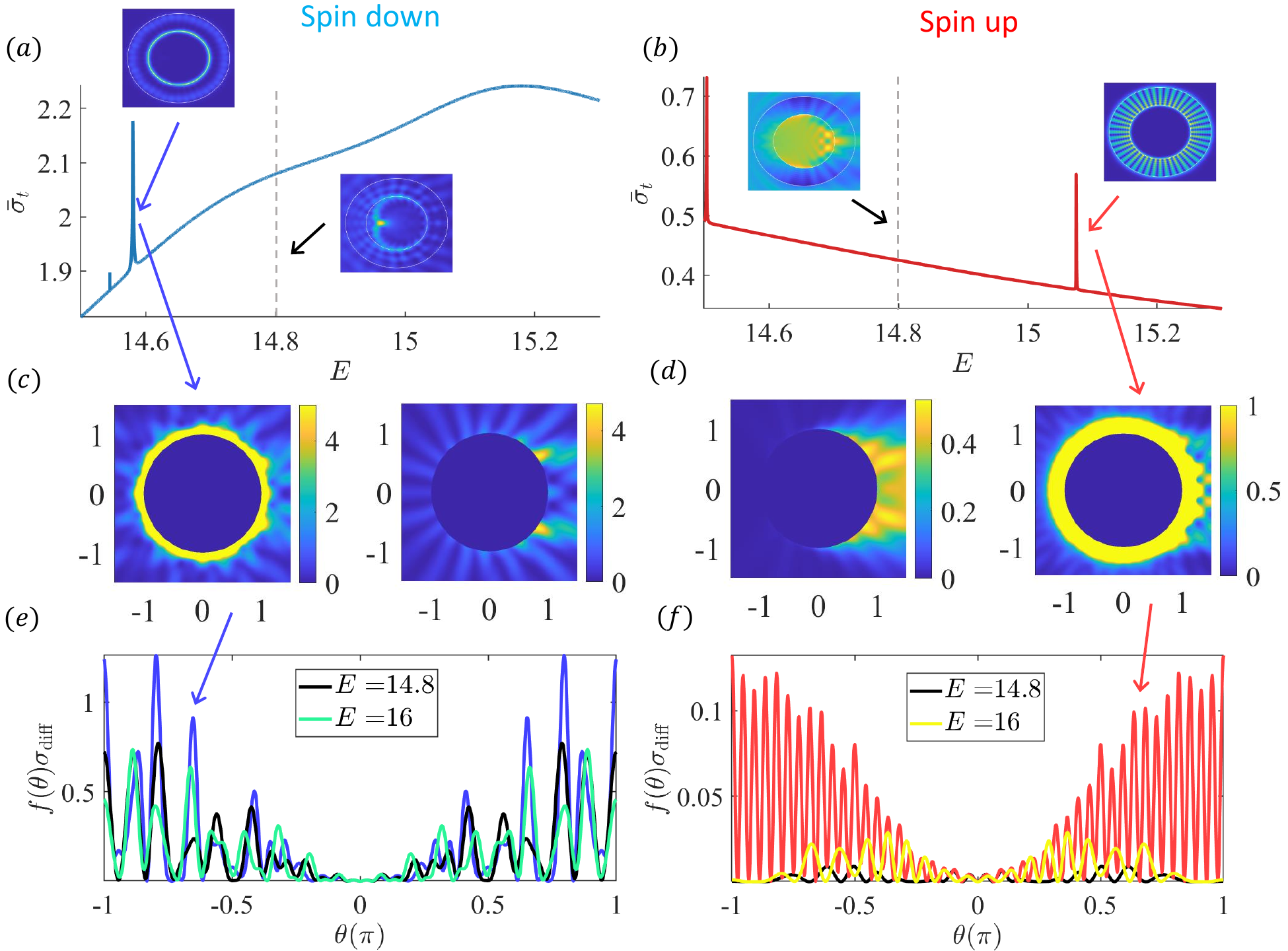}
\caption{Emergence of spin-specific edge states in an annular cavity 
($\xi = 0$). (a,b) In the reasonable Fermi energy range $\sim 0.1$ eV~\cite{tomadin:2018,ullal:2019} with $E\sim 0.01$ eV, averaged total scattering cross section, defined in Appendix~\ref{sec:crossSection}, is obtained by total scattering crossing section, $\sigma_{t}(\theta') = \oint d\theta |f(\theta,\theta')|^2$, averaging over all 
possible incident directions $\theta'$ for spin-down and spin-up electrons, 
respectively. The insets display the probability distribution patterns. 
In (a), the left inset corresponds to an edge state with no classical 
correspondence while the right inset is a conventional state. In (b), the 
scattering states are lensing-like with a classical correspondence(details in Appendix~\ref{sec:ray}). 
(c,d) In the near-field region, the scattering probability distribution defined in Appendix~\ref{sec:crossSection} in the near-field 
region $\mathrm{I}$ for spin-down and spin-up electrons, respectively, with the cut-off at the maximum value of the color bar.
(e,f) In the far-field region, differential momentum-transport cross section [differential cross section 
$\sigma_{\rm diff}$ times $f(\theta)\equiv(1-\cos\theta)$] versus the scattering 
angle $\theta$. For spin-down electrons, the total potentials are 
$V^{\downarrow}_1=-10$ and $V^{\downarrow}_2=40$. For spin-up electrons, the 
corresponding parameters are $V^{\uparrow}_1=-10-2\mu$ and $V^{\uparrow}_2=40-2\mu$
($\mu=24$). The incident plane wave is along the $x$ axis with $\theta'=0$ and the 
scattering angle is between $-\pi$ and $\pi$ (at the resolution of $1000$ points).}
\label{fig:EdgeLensing}
\end{figure*}

\begin{figure*} [ht!]
\centering
\includegraphics[width=0.7\linewidth]{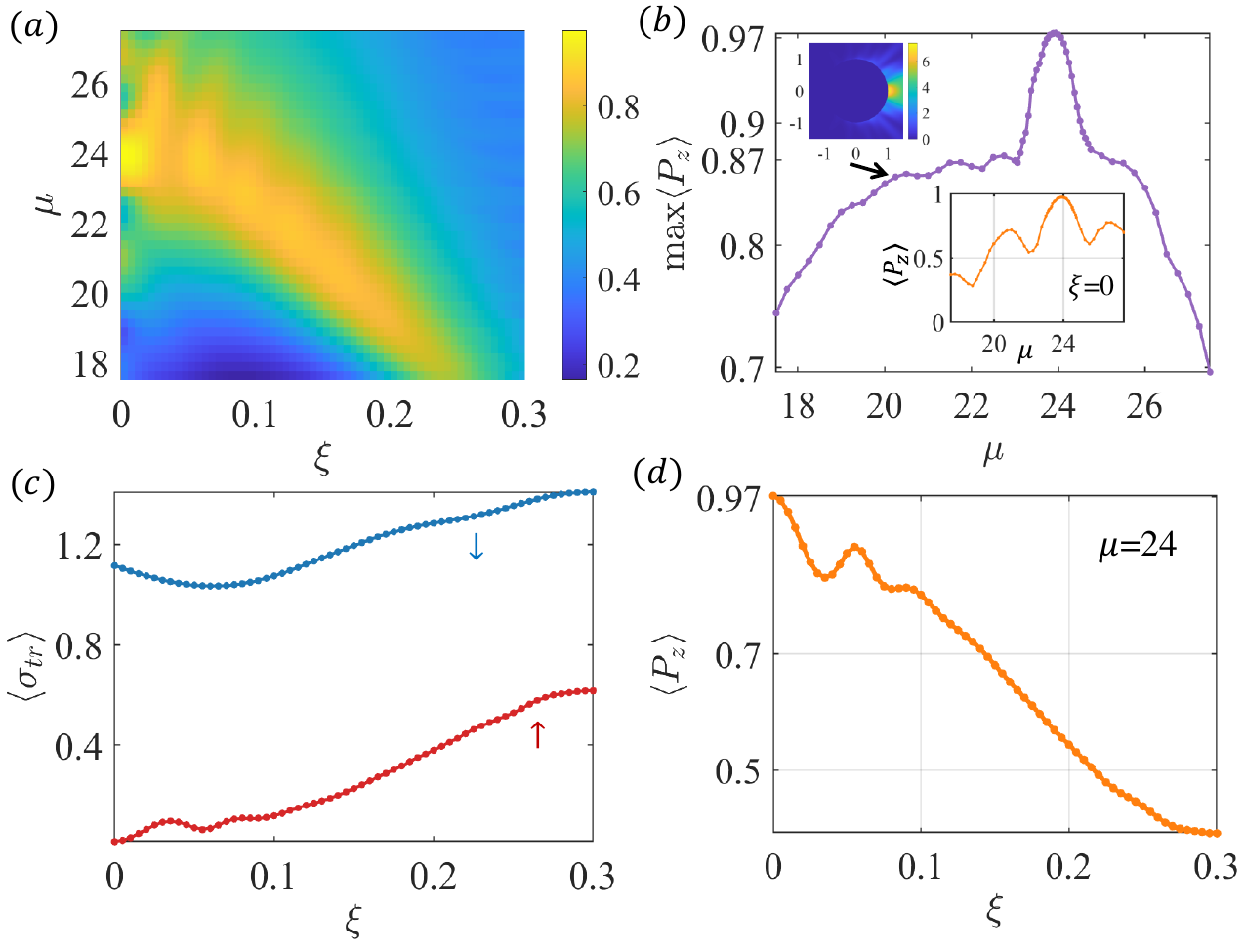}
\caption{Realization of nearly complete spin polarization. (a) Color-coded values of the 
average spin polarization $\langle P_z\rangle$ in the parameter plane $(\xi,\mu)$ 
averaged over the Fermi energy. High spin polarization can be achieved in a 
substantial area in the plane. (b) Maximum average spin polarization 
$\mathrm{max}\langle P_z\rangle$ about $\xi$ versus $\mu$. The upper inset 
displays the scattering probability density of the lensing modes for $\mu=20$, 
$E=14.8$, and $\xi=0.165$ (more details in Appendix~\ref{sec:mu_20_22}), 
and the lower inset shows $\langle P_z\rangle$ versus $\mu$ for $\xi=0$. 
Near-perfect spin polarization characterized by 
$\mathrm{max}\langle P_z\rangle \alt 1$ is achieved. (c) Momentum-transport 
cross section $\langle\sigma_{tr}\rangle$ averaged over the Fermi energy versus 
$\xi$ for the spin-down and spin-up electrons, for $\mu = 24$. 
(d) Average spin polarization $\langle P_z\rangle$ versus $\xi$ for $\mu = 24$.}
\label{fig:polar}
\end{figure*}

\section{Model}
We consider (real) spin-$1/2$ Dirac electron scattering from the 2D pseudospin-1 
Dirac system in Fig.~\ref{fig:config}(b). The eccentric circular scattering 
cavity is created by the electric gate potential $\mathcal{V}_{gate}(\mathbf{r})$~\cite{lee:2016,sadrara:2019}
and the magnetic exchange potential $\mathcal{M}(\mathbf{r})$ induced by the 
adjacent magnetic insulator within the gate region~\cite{xu:2018}. The total Hamiltonian is
\begin{align}\nonumber
\hat{H} = v_F\sigma_0\otimes\mathbf{S}\cdot\mathbf{\hat{p}} + \hbar v_F[\sigma_0\otimes &S_0 \mathcal{V}_{gate}(\mathbf{r})\\
&-\sigma_z\otimes S_0 \mathcal{M}(\mathbf{r})]
\end{align}
with pseudospin-1 matrix vector $\mathbf{S}$, spin-1/2 Pauli matrix $\sigma_z$, 
and identity matrices $\sigma_{0}^{2\times2}$ and $S_{0}^{3\times3}$. Using 
the relation $[\sigma_z\otimes S_0,\hat{H}]=0$, we block-diagonalize the 
Hamiltonian as $\hat{H}={\rm diag}[\hat{H}_1,\hat{H}_{-1}]$, where 
\begin{equation}
    \hat{H}_s = v_F\mathbf{S}\cdot\mathbf{\hat{p}}+\hbar v_F[\mathcal{V}_{gate}(\mathbf{r})-s \mathcal{M}(\mathbf{r})]
\end{equation}
for spin index $s$ ($s=1$ or $\uparrow$ for spin-up and $s=-1$ or $\downarrow$ for 
spin-down). The total potential is thus dependent upon the real spin: 
\begin{equation}
    V^{s}(\mathbf{r})\equiv\mathcal{V}_{gate}(\mathbf{r})-s \mathcal{M}(\mathbf{r}).
\end{equation}
The radii of the two eccentric circles are $R_1$ and $R_2<R_1$ whose origins are
located at $O$ and $O'$, respectively, with the eccentric distance $\xi$, as 
shown in Fig.~\ref{fig:config}(b). For $\xi \ne 0$, classical chaos can 
arise~\cite{xu:2019}. The whole physical space can be divided into three parts:
region $\mathrm{I}$ ($r>R_1$), region $\mathrm{II}$ [$r<R_1$ (origin $O$) and 
$r'>R_2$ (origin $O'$)], and region $\mathrm{III}$ ($0<r'<R_2$). The gate 
potentials $\mathcal{V}_{gate}(\mathbf{r})$ are $\nu_1$ and $\nu_2$ applied to 
regions $\mathrm{II}$ and $\mathrm{III}$, respectively, and the magnetic exchange 
potential $\mathcal{M}(\mathbf{r})$ is $\mu_1 = \mu_2 \equiv\mu$. The total 
magnetic exchange and electric potential with spin index $s$ is 
\begin{equation}
    V^s_{i}=\nu_i-s\mu
\end{equation}
for $i=1,2$ in regions $\mathrm{II}$ and $\mathrm{III}$, respectively and in the region $\mathrm{I}$, $V_0 = 0$. The energy is $\epsilon = \hbar v_F E$, where $E$ is the normalized 
energy holding the same dimension of wavelength in the unit of $1/r_0$ with the characteristic length $r_0$. The wave vectors in the three regions are
\begin{align}\nonumber
    k_{\mathrm{I}}&=|E|,\\\nonumber
    k^s_{\mathrm{II}}&=|E-V^s_1|,\\\nonumber
    k^s_{\mathrm{III}}&=|E-V^s_2|.
\end{align}
Using the principle of
Dirac electron optics~\cite{xu:2018,schrepfer:2021} and spin-resolved Snell's 
law, we have that the effective refractive indices are $n_0=(E-V_0)/E=1$ 
(vacuum) and $n^s_i=(E-V^s_{i})/E$ with $i=1,2$. 

In our work, the characteristic unit of energy, including the electronic energy, electrostatic energy, and energy of the magnetic exchange field, is $\hbar v_F/r_0\sim 0.01$ eV with $r_0=R_1\sim 100$ nm(the radius of the large circular cavity) and $v_F\sim 10^6$ m/s in 2D Dirac materials. The typical wavelength of Dirac electrons inside of the cavity is $\lambda=\hbar v_F/E_d\sim10$ nm, where $E_d$ is the energy difference between the electronic energy and the total potential with the magnitude of order $\sim 0.1$ eV. It implies the Dirac electron inside the cavity shows the particle-like behavior in the reasonable Fermi energy range $\sim 0.1$ eV~\cite{tomadin:2018,ullal:2019}, i.e. Dirac electronic optics, where the width of p-n junction edge can be efficiently sharp as $d\sim 1$ nm~\cite{balgley:2022,chen:2016,gutierrez:2016}. For convenience, the dimensions have been omitted in the following part.

\section{Results}
With the configuration in Fig.~\ref{fig:config}, the edge modes are relativistic quantum resonant states that confine the electrons 
to a quasi-1D region, where the resonant energy is about half of the 
potential. Figure~\ref{fig:EdgeLensing}(a) demonstrates an edge mode
associated with spin-down electrons (the left inset) confined around 
$r'\approx R_2$ with $E\approx15=(V_1+V_2)/2$. For comparison, the right inset 
shows a conventional pseudospin-1 scattering mode~\cite{xu:2019}.
Spin-up electrons, however, exhibit characteristically different scattering 
behaviors, as illustrated in Fig.~\ref{fig:EdgeLensing}(b) for two energy values.
The corresponding scattering probability distributions for the spin-down
and spin-up electrons are shown in Figs.~\ref{fig:EdgeLensing}(c) and 
\ref{fig:EdgeLensing}(d), respectively. The edge mode produces a large scattering
probability with wide directional distribution in both the near- and far-field
regions. [Section~\ref{subsec:chaotic_edge} provides a detailed analysis 
of the edge-mode enhanced scattering for spin-down electrons.] In contrast, the 
scattering patterns for the spin-up electrons are reminiscent of lensing 
modes in geometric optics that arise in the small wavelength limit: 
$k_{\mathrm{II}}=|E-V_1|\approx 73$, $k_{\mathrm{III}}=|E-V_2|\approx 23$, and 
$k_{\mathrm{I}}=|E|\approx 15$. The distinct scattering behaviors for spin-down
and spin-up electrons can also be characterized by the momentum-transport 
cross section, defined as $\sigma_{tr}\equiv\oint d\theta f(\theta)\sigma_{\rm diff}$
with incident direction $\theta'=0$, where $f(\theta)\equiv1-\cos\theta$, the
differential cross section $\sigma_{\rm diff}$ is determined by the scattering 
matrix, and $\sigma_{tr}$ is proportional to the resistance 
$\sigma^{\uparrow\downarrow}_{tr}\propto R^{\uparrow\downarrow}$[details in Appendix \ref{sec:crossSection}]. The edge modes 
generate a much larger resistance than the lensing states, as shown by the 
differential momentum-transport cross section in Figs.~\ref{fig:EdgeLensing}(e) 
and \ref{fig:EdgeLensing}(f), respectively.

The physical reason underlying the emergence of the edge modes lies in the 
boundary condition for the three-component spinor stipulated by the generalized 
Dirac-Weyl equation for pseudospin-1 quasiparticles~\cite{xu:2016}. In particular,
the radial or normal current density across the boundary of the scatterer must be 
continuous, but it is not necessary for the angular or tangent component of the 
current density to be continuous. In addition, the probability density needs not 
be continuous across the boundary. In fact, a larger difference in the probability
density can arise if there is a significant imbalance in the first and third
components of the spinor across the boundary. If the scattering potential 
redistributes the spinor wave-function components properly, there will be a 
significant increase in the probability density from the exterior to the interior
of the scattering boundary, leading to strong boundary trapping of the 
quasiparticles inside the potential and thereby to robust edge modes. This 
phenomenon of boundary confinement is most pronounced when the Fermi energy of the 
particle is about half of the potential height - the Klein tunneling 
regime~\cite{xu:2016}.

We now demonstrate that spin-dependent edge modes can lead to unusually nearly complete spin polarization. Figure~\ref{fig:polar}(a) shows, in the 2D parameter plane
($\xi,\mu)$, color-coded values of the spin polarization averaged over a relevant
range of the Fermi energy, which is defined as
$\langle P_z\rangle = \langle(\sigma^{\downarrow}_{tr}-\sigma^{\uparrow}_{tr})/(\sigma^{\downarrow}_{tr}+\sigma^{\uparrow}_{tr})\rangle$ from Appendix ~\ref{sec:crossSection}. 
There exists a relatively large area in the parameter plane in which the spin 
polarization exceeds $85\%$.
Figure~\ref{fig:polar}(b) shows the maximum spin polarization versus $\mu$, 
which can reach a value as high as $97\%$ (for $\mu \approx 24$), due to the
drastically different scattering behaviors associated with the spin-down and
spin-up electrons. Figure~\ref{fig:polar}(c) shows, for $\mu = 24$, the 
energy-averaged momentum-transport cross sections $\langle \sigma_{tr}\rangle$ 
versus $\xi$ for spin-down and spin-up electrons, where the cross section values 
for spin-down electrons are markedly larger than those for spin-up electrons. The 
difference is the largest for $\xi \agt 0$, leading to the highest spin 
polarization there. For a fixed value of $\mu$, as $\xi$ increases
from zero (integrable classical dynamics) to, e.g., 0.3 (chaotic classical
dynamics), the spin polarization can be maximized by some value of $\xi$ [details in Appendix \ref{sec:mu_20_22}]. Figure~\ref{fig:polar}(d) shows the average spin polarization 
versus $\xi$ for $\mu = 24$. Since $\xi$ is a geometric parameter ``controlling'' 
the degree of classical chaos (as $\xi$ increases from zero, the classical 
dynamics become more chaotic), the result shows that classical chaos deteriorates 
spin polarization.

\begin{figure} [ht!]
\centering
\includegraphics[width=\linewidth]{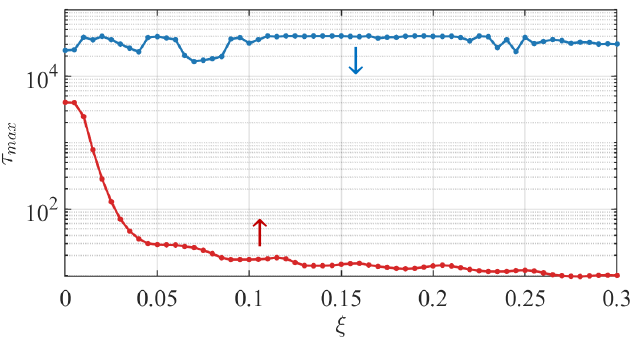}
\caption{Contrast between edge modes and lensing modes in terms of the 
Wigner-Smith time delay. Shown is a typical case of the delay maximized over the 
Fermi energy versus the geometric parameter $\xi$ associated with the edge modes 
for spin-down (blue) and the lensing modes for spin-up (red) electrons. The 
magnetic exchange potential is $\mu = 24$. The delay time for the edge modes is 
independent of the classical dynamics and is significantly longer than that for 
the lensing modes, where for the latter, the delay time decreases continuously
as the classical dynamics become more chaotic.}
\label{fig:delay}
\end{figure}
\begin{figure*} [ht!]
\centering
\includegraphics[width=0.7\linewidth]{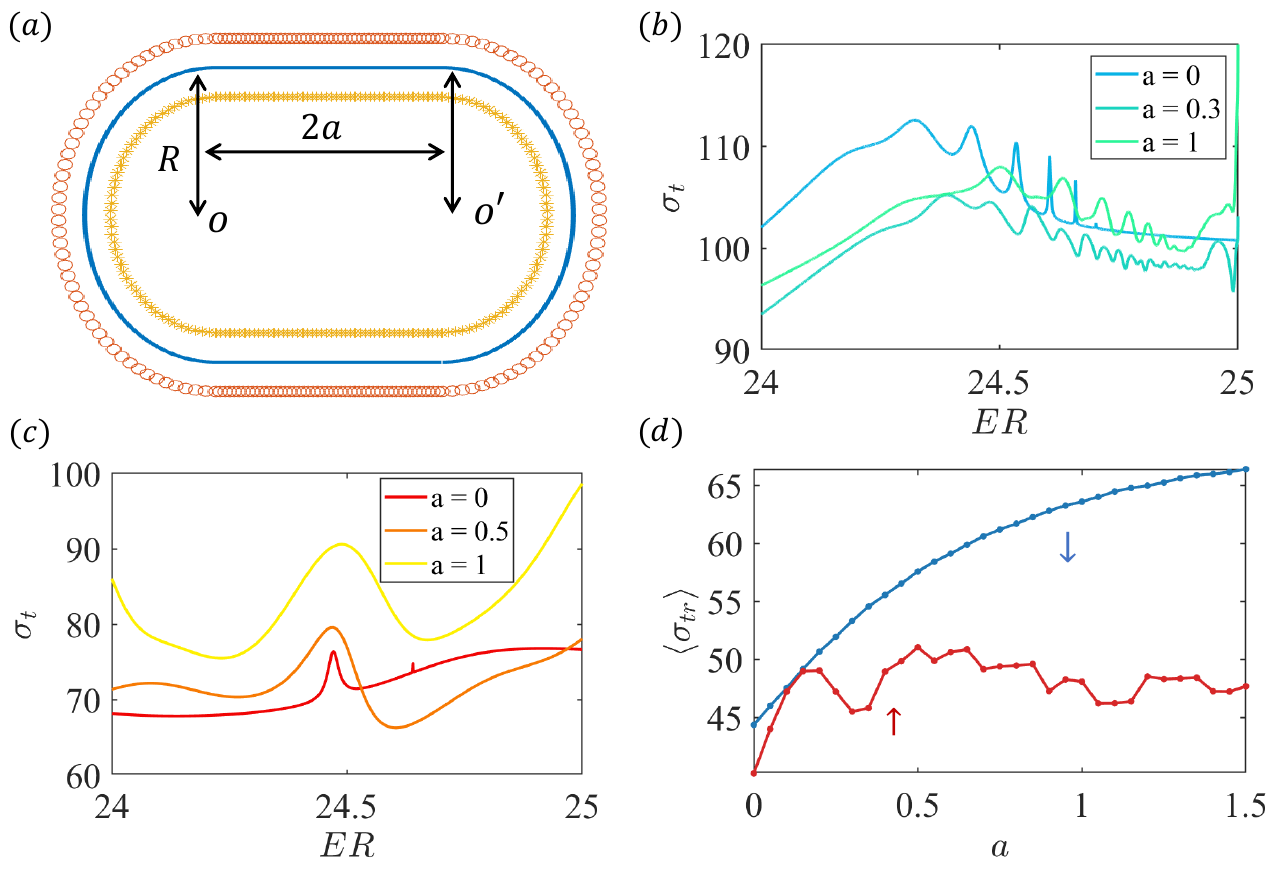}
\caption{Spin-resolved Dirac electron scattering from a stadium cavity made by the 2D pseudospin-1 Dirac material. (a) The stadium geometry (blue) defined by two parameters: ``chaotic parameter'' $a$ 
(the classical dynamics are chaotic for $a>0$) and $R$, the radius of the two 
semicircles. Two sets of ``dipoles'' are displayed, one inside and another outside 
the stadium, which are used to calculate the scattering cross sections according to 
the multiple-multipole method developed for pseudospin-1 relativistic quantum 
scattering~\cite{xu:2019}. (b) Total cross section versus the Fermi energy parameter $ER$ 
for a spin-down Dirac fermion for three values of $a$, where the total potential 
within the stadium is $V_0=\nu+\mu=50$. (c) Total cross section versus $ER$ for a 
spin-up Dirac fermion for three values of $a$, where the total potential inside the 
stadium is $V_0=\nu-\mu=70$. (d) Momentum-transport cross section 
$\langle\sigma_{tr}\rangle$ versus the chaotic parameter for a spin-down (blue) 
and a spin-up (red) electron.}
\label{fig:suppStadiumCase}
\end{figure*}

The characteristic difference between the edge modes for spin-down electrons and 
the lensing modes for spin-up electrons can also be revealed by the maximum 
Wigner-Smith time delay defined as 
$\tau(E)\equiv-i\hbar \mathrm{Tr}[S^{\dagger}\partial S/\partial E]$, 
with $S$ being the scattering matrix in Appendix~\ref{sec:crossSection}. Figure~\ref{fig:delay} shows,
for $\mu = 24$, the maximum delay $\tau_{max}$ (over Fermi energy) versus the geometric 
parameter $\xi$ for spin-down (blue) and spin-up (red) electrons, where the former
is significantly larger than that for the latter. A remarkable feature is that,
as $\xi$ increases from zero so that the classical dynamics changes from being
integrable to mixed and then to chaotic, $\tau_{max}$ for spin-down electrons
hardly vary, indicating that the edge modes have no classical correspondence.
In contrast, $\tau_{max}$ for spin-up electrons continue to decrease with $\xi$,
which agrees with the classical intuition that, as the dynamics become more 
chaotic, the average time that an electron can stay inside the cavity should 
decrease. Because of the classical-quantum correspondence for the lensing modes,
their properties can be understood using ray tracing from geometric 
optics in Appendix~\ref{sec:ray}.

\section{Discussion}

\subsection{Emergence of edge mode in the chaotic stadium cavity}\label{subsec:chaotic_edge}

In a previous work~\cite{xu:2019}, it was demonstrated that an edge mode can 
confine a particle for a long time, defying any Q-spoiling effect induced by 
classical chaos. To further demonstrate the ``peculiar'' behavior of the edge modes,
we set up and study a spin-resolved scattering cavity of the stadium shape, whose
geometric boundary is shown as the blue curve in Fig.~\ref{fig:suppStadiumCase}(a),
where $a$ is a so-called ``chaotic parameter'' in the sense that the classical
dynamics are chaotic for $a>0$. To calculate the scattering cross sections,
we use a previously developed method, the multiple-multipole method~\cite{xu:2019}, 
where two sets of ``dipoles'', one inside and another outside the cavity, as
shown in Fig.~\ref{fig:suppStadiumCase}(a), are used as the sources to produce the 
far-field scattering wave function. For spin-down electrons, the total potential
in the cavity is $V_0=50$. There are quasibound edge modes with Fermi energy about half 
of the total potential, as shown by the peaks in the total cross section in 
Fig.~\ref{fig:suppStadiumCase}(b). For spin-up electrons, the total potential in 
the cavity is $V_0=70$ and the classical dynamics are chaotic, which 
smooths out the sharp resonances, as shown in Fig.~\ref{fig:suppStadiumCase}(c). 
For the edge mode associated with spin-down electrons, the resonant peaks have 
also been smoothed out. Intuitively, a larger potential in the cavity
produces stronger scattering. However, the edge mode leads to strong scattering
even with a small potential, as shown in Fig.~\ref{fig:suppStadiumCase}(d), the
momentum-transport cross section versus the stadium parameter $a$. It can be seen
that a spin-down electron, due to its large momentum-transport cross section
$\langle\sigma_{tr}\rangle$ as the result of the edge mode, produces larger and 
larger equivalent scattering resistance than that from a spin-up electron as the 
chaotic parameter increases.

\subsection{Conclusion and outlook}
We design the whole scatterer placed on a magnetic insulator substrate so that the real spin degree of the electrons matters in the sense that the spin-up and spin-down electrons will 
experience a different magnetic exchange potential. Based on this, we articulated a simple eccentric circular scatterer to generate edge modes for electrons with a specific spin orientation, where the electrons can be confined around the edge modes for a long time, generating resonant scattering with a large momentum-transport cross section. The quantum scattering behaviors for these electrons do not have a classical correspondence. On the contrary, electrons with the opposite spin will not possess such
edge modes: they tend to stay in the scattering region for a much shorter time with a small cross section. For these electrons, the quantum scattering dynamics have a classical correspondence, so ray tracing with Dirac electron optics can be used to understand their behaviors (see Appendix~\ref{sec:ray}). The remarkable difference in the 
spin-specific scattering cross sections leads to tunable spin polarization and can even generate near-perfect spin polarization. The physical principles laid out in this work are anticipated to find applications in spintronics.

The basic principle of spintronics is to manipulate the spin degree of freedom
to bring new capabilities to microelectronics and information technology with 
applications such as magnetic memories and sensors, radio-frequency
and microwave devices, and logic and non-Boolean devices~\cite{vzutic:2004}.
In spintronics, a key requirement is to achieve high spin polarization in 
functional materials~\cite{dieny:2020}, which has remained to be a challenge. 
For example, the early proposition of spin field-effect transistors for 
large-scale integrated circuits~\cite{datta:1990} requires high spin 
polarization~\cite{chuang:2015,yan:2016,jiang:2019,dieny:2020,malik:2020,liu:2021}.
Graphene spintronics~\cite{han:2014} based on relativistic quantum mechanics of 
pseudospin-1/2 fermions possess certain advantages such as room-temperature spin 
transport with long spin diffusion lengths of several 
micrometers~\cite{tombros:2007,yang:2011}, gate-tunable carrier concentration, 
high electronic mobility, and efficient spin injection~\cite{han:2010,dlubak:2012}. 
However, even for graphene, designing a system configuration to achieve high
spin polarization is challenging~\cite{xu:2018} but holds some breakthroughs. For instance, the work in~\cite{maksym:2021} realizes 100\% spin and valley polarized in monolayer transition metal dichalcogenides(TMD) assisted by total external reflection with spin-orbit coupling and electrostatic potential barrier. Another work~\cite{zeng:2011} also realizes the nearly ±100\% spin-polarized current by the magnetic configuration in two-terminal bipolar spin diodes of zigzag graphene nanoribbons. Although the artificial magnetic field in 2D materials can also produce spin polarization and other intriguing physical effects, it requires the systematical technology to employ and control the magnetic field, of which the magnitude is hard to achieve the order of tesla. Conversely, adding magnetic insulators~\cite{wei:2016,singh:2017} or magnetic impurities~\cite{dugaev:2006} on top of 2D materials can induce the magnetic exchange field (MEF), which can potentially reach at least several tesla magnitudes~\cite{li:2013,wei:2016}. In addition, MEF facilitates extensive research on electronic optics~\cite{grivet:2013,batson:2002,chen:2016,tian:2012,xu:2018,wang:2019chaos,schrepfer:2021,wang:2022}. To summarize, it holds the fundamental and applicable interest to explore the physical nature of 2D materials with MEF interaction.

Experimentally, it has become feasible to implement electron scattering in 2D 
Dirac materials. For example, the width of p-n junction edge in Dirac materials 
can already be made sufficiently sharp~\cite{balgley:2022,chen:2016,gutierrez:2016}
(e.g., $d\sim 1$ nm compared with the typical Fermi wavelength 
$\lambda_F\sim 10$ nm). In addition, the materials can be fabricated on the 
scale of micrometers to reach the small wavelength limit at which Dirac electron 
optics is applicable~\cite{jiang:2017}. The required magnetic exchange potential 
has been realized in experiments~\cite{wei:2016,singh:2017}. For electrostatic potential in the eccentric circular shape, in the recent experiment~\cite{lee:2016}, a circular p-n junction,  i.e. a local embedded gate, in a graphene/hBN heterostructure is created by local defect charge and STM tip with a square voltage pulse. Moreover, the Dirac electron scattering in the multi-circular quantum dots has been discussed~\cite{sadrara:2019}. It implies the possibility of fabrication of the eccentric circular cavity shape by STM technology. Experimental 
material platforms have also existed to create pseudospin-1 Dirac systems with a 
flat band, such as the transition-metal oxide $\mbox{SrTiO}_3/\mbox{SrIrO}_3/\mbox{SrTiO}_3$ trilayer heterostructures~\cite{Wang:2011}, 
$\mbox{SrCu}_{2}(\mbox{BO}_{3})_{2}$~\cite{romhanyi:2015}, and 
graphene-$\mbox{In}_{2}\mbox{Te}_{2}$ bilayer~\cite{giovannetti:2015}.

\section*{Acknowledgment}

This work was supported by AFOSR under Grant No.~FA9550-21-1-0186.

\appendix
\section{S-Matrix approach to elastic Dirac electron scattering}

Consider electronic scattering from a cavity made of two-dimensional (2D) Dirac 
materials with a flat band. At low energies, the effective Hamiltonian describes 
the dynamics of a pseudospin-1 Dirac-Weyl quasiparticle. The cavity is subject to
external electrical and magnetic exchange fields: its properties are controlled by an 
electric gate potential $\mathcal{V}_{gate}(\mathbf{r})$ and a magnetic exchange 
potential $\mathcal{M}(\mathbf{r})$ induced by the magnetic insulator substrate within the gate 
region~\cite{xu:2018}. The total Hamiltonian is 
\begin{align} \label{eq:H}\nonumber
\hat{H} = v_F\sigma_0\otimes\mathbf{S}\cdot\mathbf{\hat{p}}+\hbar v_F[\sigma_0\otimes &S_0 \mathcal{V}_{gate}(\mathbf{r})\\
&-\sigma_z\otimes S_0 \mathcal{M}(\mathbf{r})],
\end{align}
where $\mathbf{S}$ denotes the vector of spin-1 matrices, $\sigma_0$ and $S_0$ 
are the two-by-two and three-by-three identity matrices, respectively, $\sigma_z$ 
is the Pauli $z$ matrix, and $v_F$ is the Fermi velocity. Tensor product of the three-component pseudospin-1 quasiparticles and two-component real spin $1/2$ electron, so the Hamiltonian matrix is six-by-six, which
can be block-diagonalized as $\hat{H}={\rm diag}[\hat{H}_1,\hat{H}_{-1}]$ with the 
following two three-by-three sub-Hamiltonian matrices $\hat{H}_s$ for real spin 
index $s=\pm1$: 
\begin{align} \label{eq:H_s}
    \hat{H}_s = v_F\mathbf{S}\cdot\mathbf{\hat{p}}+\hbar v_F[\mathcal{V}_{gate}(\mathbf{r})-s \mathcal{M}(\mathbf{r})],
\end{align}
where the identity $[\sigma_z\otimes S_0,\hat{H}]=0$ has been used. The total 
potential is spin-dependent: 
\begin{align} \nonumber
V^{s}(\mathbf{r})\equiv\mathcal{V}_{gate}(\mathbf{r})-s\mathcal{M}(\mathbf{r}). 
\end{align}
The prototypical system we use to demonstrate achieving high spin 
polarization is an eccentric circular cavity defined by two distinct radii: 
$R_1$ and $R_2<R_1$, where the centers of the two circles are located at $O$ (the 
larger disk) and $O'$ (the smaller disk) with the eccentric distance $\xi$ between 
$OO'$, as shown in Fig.~1(b) in the main text. For convenience, we define three 
regions in the position space: region $\mathrm{I}$ with $V_0=0$ for $r>R_1$, region 
$\mathrm{II}$ with $V^{s}_1$ for $r<R_1$ and $r'>R_2$, and region $\mathrm{III}$ 
with $V^{s}_2$ for $r'<R_2$. The wave vectors in the three regions are given by
\begin{align} \nonumber
	k_{\mathrm{I}} &= |E|, \\ \nonumber
	k^{s}_{\mathrm{II}} &=|E-V^{s}_{\mathrm{1}}|, \\ \nonumber
	k^{s}_{\mathrm{III}} &=|E-V^{s}_{\mathrm{2}}|.
\end{align}
The wave functions in the three regions can be written down according to the
standard form of the spinor wave eigenvector of $\hat{H}_s$ in the cylindrical 
coordinates, which are given by
\begin{align}\label{eq:general_basis}
    ^{k}g_m=\frac{1}{\sqrt{2}} \left(\begin{array}{c}f_{m-1}(kr)e^{-i\theta} \\
i\alpha\sqrt{2}f_{m}(kr) \\
-f_{m+1}(kr)e^{i\theta} \end{array}\right)e^{i m\theta},
\end{align}
where $\alpha \equiv \mathrm{sign}(E-V)$, $k=|E-V|$. There are two cases for the
function $f_m(kr)$: (i) $f_m=H^{(1,2)}_m$, the Hankel functions of the first and 
the second kind, and (ii) $f_m=J_m$, the Bessel function. For cases (i) and (ii),  
$^{k}g_m$ is given by ${^{k}g_m}={^{k}h^{(1,2)}_m}$ and ${^{k}g_m}={^{k}j_m}$, 
respectively. In particular, in region $\mathrm{I}$, the wave function can be 
expanded in the spinor cylindrical wave basis as
\begin{align} \label{eq:psi1}
    \Psi^{(\mathrm{I})}(\Vec{r})=\sum_{m=-\infty}^{+\infty}a^{\mathrm{I}}_m\left[{^{k_\mathrm{I}}h^{(2)}_m}+\sum_{j=-\infty}^{+\infty}S_{mj}\;{^{k_\mathrm{I}}h^{(1)}_{j}}\right].
\end{align}
In region II, the wave function can be written as
\begin{align}
     \Psi^{(\mathrm{II})}(\Vec{r})=\sum_{m=-\infty}^{+\infty}\sum_{l=-\infty}^{+\infty}{^m}a^{\mathrm{II}}_{l}\left[{^{k_{\mathrm{II}}}h^{(2)}_{l}}+\sum_{j=-\infty}^{+\infty}S^{od}_{lj}\;{^{k_{\mathrm{II}}}h^{(1)}_{j}}\right],
\end{align}
where $S^{od}$ is the off-diagonal scattering matrix for the eccentric circular 
cavity and $S^{cd}$ is the diagonal matrix to characterize the scattering from a 
circular domain~\cite{xu:2018}, which are related by $S^{od}=U^{-1}S^{cd}U$, or
\begin{align} \nonumber
    S^{od}_{lj}&=\sum_{l',j'}(U^{-1})_{ll'}S^{cd}_{l'j'}U_{j'j}\\\nonumber
    &=\sum_{l',j'}J_{l-l'}(k_{\mathrm{II}}\xi)S^{cd}_{l'l'}\delta_{l'j'}J_{j-j'}(k_{\mathrm{II}}\xi)\\\nonumber
    & = \sum_{l'}J_{l-l'}S^{cd}_{l'l'}J_{j-l'}.
\end{align}
The boundary conditions for a pseudospin-1 quasiparticle~\cite{xu:2016} stipulate
continuity of the second component of the spinor wave function and conservation of 
the radial current density:
\begin{align} \label{eq:boundaryCondition}
    \Psi^{\mathrm{I}}_2(R_1)&=\Psi^{\mathrm{II}}_2(R_1), \\
    \Psi^{\mathrm{I}}_1(R_1)e^{i\theta}+\Psi^{\mathrm{I}}_3(R_1)e^{-i\theta}&=\Psi^{\mathrm{II}}_1(R_1)e^{i\theta}+\Psi^{\mathrm{II}}_3(R_1)e^{-i\theta}.
\end{align}
\begin{align}\nonumber
    \Psi^{\mathrm{I}}_1(R_1)&=\Psi^{\mathrm{II}}_1(R_1), \\ \nonumber
     \Psi^{\mathrm{I}}_2(R_1)&=\Psi^{\mathrm{II}}_2(R_1).
\end{align}
In matrix form, the boundary conditions can be expressed as
\begin{align}
    A^{\mathrm{I}}[X^{(2)}+SX^{(1)}]&=\alpha_{\mathrm{I}}\alpha_{\mathrm{II}}A^{\mathrm{II}}[x^{(2)}+S^{od}x^{(1)}],\\
    A^{\mathrm{I}}[[Z^{(2)}-Y^{(2)}]&+S[Z^{(1)}-Y^{(1)}]]\\\nonumber
    &=A^{\mathrm{II}}[[z^{(2)}-y^{(2)}]+S^{od}[z^{(1)}-y^{(1)}]],
\end{align}
where $\alpha_{\mathrm{I}}\equiv\mathrm{sign}(E)$, 
$\alpha_{\mathrm{II}}\equiv\mathrm{sign}(E-V_1)$, 
$A^{\mathrm{I}}\equiv[a^{\mathrm{I}}_{m}\delta_{mj}]$, 
$A^{\mathrm{II}}\equiv[{^m}a^{\mathrm{II}}_{j}]$ and
\begin{align}\nonumber
    X^{(1,2)}&\equiv[H^{(1,2)}_m(k_{\mathrm{I}}R_1)\delta_{mj}],\; x^{(1,2)}\equiv[H^{(1,2)}_m(k_{\mathrm{II}}R_1)\delta_{mj}];\\ \nonumber
    Y^{(1,2)}&\equiv[H^{(1,2)}_{m+1}(k_{\mathrm{I}}R_1)\delta_{mj}],\; y^{(1,2)}\equiv[H^{(1,2)}_{m+1}(k_{\mathrm{II}}R_1)\delta_{mj}];\\
    Z^{(1,2)}&\equiv[H^{(1,2)}_{m-1}(k_{\mathrm{I}}R_1)\delta_{mj}],\; z^{(1,2)}\equiv[H^{(1,2)}_{m-1}(k_{\mathrm{II}}R_1)\delta_{mj}].
\end{align}
The spinor wave function can be written as
\begin{align}
    ^{k}h_m^{(1,2)}=\frac{1}{\sqrt{2}} \left(\begin{array}{c}H^{(1,2)}_{m-1}(kr)e^{-i\theta} \\
i\alpha\sqrt{2}H^{(1,2)}_{m}(kr) \\
-H^{(1,2)}_{m+1}(kr)e^{i\theta} \end{array}\right)e^{i m\theta},
\end{align}
where the general form of the basis is described by Eq.~\eqref{eq:general_basis}.
The scattering matrix can be written as
\begin{equation} \label{eq:SmatrixEccentric}
    S = -\frac{Z^{(2)}-Y^{(2)}-\alpha_{\mathrm{I}}\alpha_{\mathrm{II}}X^{(2)}\mathcal{T}}{Z^{(1)}-Y^{(1)}-\alpha_{\mathrm{I}}\alpha_{\mathrm{II}}X^{(1)}\mathcal{T}},
\end{equation}
where $\mathcal{T}\equiv F^{-1}[H-G]$, and
\begin{align}\nonumber
    F&\equiv x^{(2)}+S^{od}x^{(1)},\\\nonumber
    G&\equiv y^{(2)}+S^{od}y^{(1)},\\
    H&\equiv z^{(2)}+S^{od}z^{(1)}.
\end{align}
The coefficient $A^{\mathrm{I}}$ is determined by the incident wave function (see 
Appendix~\ref{sec:crossSection}), and the coefficient $A^{\mathrm{II}}$ is given by
\begin{align} \nonumber
A^{\mathrm{II}}=\alpha_{\mathrm{I}}\alpha_{\mathrm{II}}A^{\mathrm{I}}[X^{(2)}+SX^{(1)}]F^{-1}.
\end{align}
Using the Graf's addition theorem~\cite{sadrara:2019}, we have, for $r'>\xi$,
\begin{equation}
    H^{(1,2)}_m(kr)e^{im\theta}=\sum_{n=-\infty}^{+\infty}J_{m-n}(k\xi)e^{in\theta'}H^{(1,2)}_{n}(kr'),
\end{equation}
which gives
\begin{equation}
    {^{k}}h_m^{(1,2)} = \sum_{n=-\infty}^{+\infty}J_{m-n}(k\xi)\;{^{k}}\widetilde{h}_n^{(1,2)}.
\end{equation}
For convenience, in the following, we use the tilde symbol to denote the quantities
in the circular region of origin at $O'$. We have
\begin{widetext}
\begin{align}\nonumber
    {^{k_{\mathrm{II}}}h^{(2)}_{l}}+\sum_{j=-\infty}^{+\infty}S^{od}_{lj}\;{^{k_{\mathrm{II}}}h^{(1)}_{j}}&=\sum_{n=-\infty}^{+\infty}J_{l-n}(k_{\mathrm{II}}\xi)\;{^{k_{\mathrm{II}}}}\widetilde{h}^{(2)}_n+\sum_{j=-\infty}^{+\infty}S^{od}_{lj}\left[\sum_{n=-\infty}^{+\infty}J_{j-n}(k_{\mathrm{II}}\xi)\;{^{k_{\mathrm{II}}}}\widetilde{h}^{(1)}_n\right]\\\nonumber
    &=\sum_{l'=-\infty}^{+\infty}J_{l-l'}\;{^{k_{\mathrm{II}}}}\widetilde{h}^{(2)}_{l'}+\sum_{n,l'=-\infty}^{+\infty}J_{l-l'}S^{cd}_{l'l'}\sum_{j=-\infty}^{+\infty}(J_{j-l'}J_{j-n})\;{^{k_{\mathrm{II}}}}\widetilde{h}^{(1)}_n\\
    &=\sum_{l'=-\infty}^{+\infty}J_{l-l'}(k_{\mathrm{II}}\xi)\left[{^{k_{\mathrm{II}}}\widetilde{h}^{(2)}_{l'}}+S^{cd}_{l'l'}\;{^{k_{\mathrm{II}}}\widetilde{h}^{(1)}_{l'}}\right],
\end{align}
\end{widetext}
where 
\begin{align} \nonumber
\delta_{l'n}=\sum_{j=-\infty}^{+\infty}J_{j-l'}(k_{\mathrm{II}}\xi)J_{j-n}(k_{\mathrm{II}}\xi).
\end{align}
The wave function in region $\mathrm{II}$ with origin $O$ can be rewritten as a 
wave function with origin $O'$ as 
$\Psi^{\mathrm{II}}(r,\theta)=\widetilde{\Psi}^{\mathrm{II}}(r',\theta')$, where
\begin{align}\nonumber
    \widetilde{\Psi}^{\mathrm{II}}(r',\theta')=\sum_{m=-\infty}^{+\infty}\sum_{l=-\infty}^{+\infty}{^m}\widetilde{a}^{\mathrm{II}}_l\left[{^{k_{\mathrm{II}}}}\widetilde{h}^{(2)}_l+S^{cd}_{ll}\;{^{k_{\mathrm{II}}}}\widetilde{h}^{(1)}_l\right],
\end{align}
with ${^m}\widetilde{a}^{\mathrm{II}}_l\equiv\sum_{l'}{^m}a^{\mathrm{II}}_{l'}J_{l'-l}(k_{\mathrm{II}}\xi)$. 
In region $\mathrm{III}$ with origin $O'$, the wavefunction is given by
\begin{equation}
    \widetilde{\Psi}^{\mathrm{III}}(r',\theta')=\sum_{m=-\infty}^{+\infty}\sum_{l=-\infty}^{+\infty}{^m}\widetilde{b}_l\;{^{k_{\mathrm{III}}}}\widetilde{j}_{l}.
\end{equation}
Using the boundary condition Eq.~(\ref{eq:boundaryCondition}), we obtain
\begin{equation}
    {^m}\widetilde{b}_l=\alpha_{\mathrm{II}}\alpha_{\mathrm{III}}{^m}\widetilde{a}^{\mathrm{II}}_l\;\frac{H^{(2)}_l(k_{\mathrm{II}}R_2)+S^{cd}_{ll}H^{(1)}_{l}(k_{\mathrm{II}}R_2)}{J_{l}(k_{\mathrm{III}}R_2)}.
\end{equation}

\section{Scattering matrix for a circular cavity}

To obtain the scattering matrix $S^{cd}$, we consider a circular cavity of radius 
$R_2$ centered at $O$ where $r>R_2$ and $0<r<R_2$ define regions $\mathrm{II}$ and
$\mathrm{III}$, respectively. Due to the circular symmetry, the wave function for 
each angular momentum channel can be written as 
\begin{align}\nonumber
    \Psi^{\mathrm{II}}_{m} &= {^{k_{\mathrm{II}}}}h^{(2)}_m+S^{cd}_{mm}\;{^{k_{\mathrm{II}}}}h^{(1)}_{m}\\
    \Psi^{\mathrm{III}}_m &=B_{m}{^{k_{\mathrm{III}}}}j_{m}.
\end{align}
Applying the boundary conditions gives 
\begin{align}\nonumber
    \alpha_{\mathrm{II}}\left[H^{(2)}_{m}(k_{\mathrm{II}}R_2)+S^{cd}_{mm}H^{(1)}_{m}(k_{\mathrm{II}}R_2)\right]&=\alpha_{\mathrm{III}}B_{m}J_{m}(k_{\mathrm{III}}R_2)\\
    {^{k_{\mathrm{II}}}_{1+3}}h^{(2)}_{m}(R_2)+S^{cd}_{mm}~{^{k_{\mathrm{II}}}_{1+3}}h^{(1)}_{m}(R_2)&=B_m ~{^{k_{\mathrm{III}}}_{1+3}}j_{m}(R_2),
\end{align}
where 
\begin{align}
{^{k_{\mathrm{II}}}_{1+3}}h^{(1,2)}_{m}(R_2)&\equiv H^{(1,2)}_{m-1}(k_{\mathrm{II}}R_2)-H^{(1,2)}_{m+1}(k_{\mathrm{II}}R_2) \\
{^{k_{\mathrm{III}}}_{1+3}}j_{m}(R_2)&\equiv J_{m-1}(k_{\mathrm{III}}R_2)-J_{m+1}(k_{\mathrm{III}}R_2).
\end{align}
We thus have
\begin{widetext}
    \begin{equation}
    S^{cd}_{mm}=-\frac{J_{m}(k_{\mathrm{III}}R_2)~{^{k_{\mathrm{II}}}_{1+3}}h^{(2)}_{m}(R_2)-\alpha_{\mathrm{II}}\alpha_{\mathrm{III}}H^{(2)}_m(k_{\mathrm{II}}R_2)~{^{k_{\mathrm{III}}}_{1+3}}j_{m}(R_2)}{J_{m}(k_{\mathrm{III}}R_2)~{^{k_{\mathrm{II}}}_{1+3}}h^{(1)}_{m}(R_2)-\alpha_{\mathrm{II}}\alpha_{\mathrm{III}}H^{(1)}_m(k_{\mathrm{II}}R_2)~{^{k_{\mathrm{III}}}_{1+3}}j_{m}(R_2)}~,
\end{equation}
\end{widetext}
with $\alpha_{\mathrm{III}}\equiv\mathrm{sign}(E-V_2)$.

\section{Scattering cross sections} \label{sec:crossSection}

The wave function in region $\mathrm{I}$ from Eq.~(\ref{eq:psi1}) can be rewritten 
as the sum of the contributions from the incident and scattering waves:
\begin{align} \label{eq:psi1Decompose}
    \Psi^{(\mathrm{I})}(\Vec{r})&=\sum_{m=-\infty}^{\infty}a^{\mathrm{I}}_m\left[2\;{^{k_\mathrm{I}}j_m}+\sum_{m'=-\infty}^{\infty}(S_{mm'}-\delta_{mm'}){^{k_\mathrm{I}}h^{(1)}_{m'}}\right]\nonumber\\
     &=\chi_{in}+\sum_{m=-\infty}^{\infty}a^{\mathrm{I}}_m\sum_{m'=-\infty}^{\infty}T_{mm'}~{^{k_\mathrm{I}}h^{(1)}_{m'}}.
\end{align}
The incident wave function corresponds to 
\begin{align} \label{eq:incident}
\chi_{in}\equiv\sum_{m=-\infty}^{\infty}2a^{\mathrm{I}}_{m}{^{k_{\mathrm{I}}}j_m}.
\end{align}
The norm square of the second term in Eq.~\eqref{eq:psi1Decompose}, which is the 
scattering wave function, is defined as the scattering probability in the near field
measured from the cavity in region $\mathrm{I}$, with the transmission matrix 
defined as $T_{mm'}\equiv S_{mm'} - \delta_{mm'}$. The coefficient 
$a^{\mathrm{I}}_m$ for each angular momentum channel is determined by the incident 
plane wave function:
\begin{align} \label{eq:chi_in}
    \chi_{in}(r,\theta)&=\frac{1}{2}\left(\begin{array}{c}
    e^{-i\theta'}\\
    \sqrt{2}s\\
     e^{i\theta'}
    \end{array}\right)e^{i\mathbf{k}_{in}\cdot\mathbf{r}},
\end{align}
with the incident wave vector 
$\mathbf{k}_{in}=k_{\mathrm{I}}(\cos\theta',\sin\theta')$. Expanding the incident 
wave function for each angular momentum channel by the Jacobi-Anger formula
\begin{equation}
    e^{ik_{\mathrm{I}}r\cos(\theta-\theta')}=\sum_{m=-\infty}^{+\infty}i^{m}J_{m}(k_{\mathrm{I}}r)e^{im(\theta-\theta')},
\end{equation}
we obtain
\begin{equation}
    \chi_{in}(r,\theta)=\sum_{m=-\infty}^{+\infty}\frac{1}{\sqrt{2}}\;i^{m-1}e^{-im\theta'}\;{^{k_{\mathrm{I}}}}j_m.
\end{equation}
Note that $j_m$ is the three components vector defined by 
Eq.~\eqref{eq:general_basis} while $J_m$ is the scalar Bessel function. The 
coefficient $a^{\mathrm{I}}_m(\theta')$ in Eq.~\eqref{eq:psi1Decompose} and 
Eq.~\eqref{eq:incident} is then given by
\begin{align} \label{eq:coefficient_a}
a^{\mathrm{I}}_m(\theta')=i^{m-1}e^{-im\theta'}/(2\sqrt{2}).
\end{align}
Scattering cross section characterizes the behavior of particles in the far-field 
region $kr\gg 1$ (from the cavity). In the far field, the wave function from 
Eq.~\eqref{eq:psi1Decompose} tends to
\begin{equation}\label{eq:boundary_infinite}
    \lim_{kr\gg 1}\psi(r,\theta) = \chi_{in} + \frac{f(\theta,\theta')}{2\sqrt{-ir}}\left(\begin{array}{c}
        e^{-i\theta} \\
        \sqrt{2}s\\
        e^{i\theta}
    \end{array}\right)e^{ik_{\mathrm{I}}r}
\end{equation}
with the scattering angle distribution in the far field as
\begin{align} \label{eq:f_theta}
f(\theta,\theta') = \frac{2}{\sqrt{\pi k_{\mathrm{I}}}}\sum_{m=-\infty}^{\infty}a_{m}^{\mathrm{I}}(\theta')\sum_{m'=-\infty}^{\infty}T_{mm'}(-i)^{m'}e^{im'\theta},
\end{align}
a result of the asymptotic behavior of the Hankel function: 
\begin{align} \nonumber
\lim_{x\rightarrow{}\infty} H^{(1,2)}_{m}(x)\rightarrow \sqrt{\frac{2}{\pi x}}e^{\pm i(x-m\pi/2-\pi/4)}, 
\end{align} 
and the standard plane-wave normalization requirement. The differential cross 
section  $\sigma_{\rm diff}$ is given in terms of $f(\theta,\theta')$ as
\begin{equation}
    \sigma_{\rm diff}\equiv\frac{d\sigma}{d\theta} = |f(\theta,\theta')|^2,
\end{equation}
and the total scattering cross section, which records the probability of scattering 
events under all possible directions, is given by 
\begin{equation}
    \sigma_{t}(\theta') = \oint d\theta |f(\theta,\theta')|^2.
\end{equation}
The momentum-transport cross section is defined as
\begin{equation}
    \sigma_{tr}(\theta') = \oint d \theta (1-\cos\theta) |f(\theta,\theta')|^2.
\end{equation}
Averaging over the incident angle $\theta'$ leads to
\begin{align}
\bar{\sigma}_t = \frac{1}{2\pi}\oint d\theta' \sigma_{t}(\theta'), \\ 
\bar{\sigma}_{tr} = \frac{1}{2\pi}\oint d\theta' \sigma_{tr}(\theta').
\end{align}
Performing an average over some Fermi energy interval, we get
\begin{align}
    \langle\sigma_{tr}\rangle = \frac{1}{E_1-E_0}\int^{E_1}_{E_0} dE ~\sigma_{tr}(E).
\end{align}
The momentum transport cross section determines the transport relaxation time $\tau_{tr}$ through
\begin{equation}
    \frac{1}{\tau_{tr}}=n_c v_{F}\sigma_{tr},
\end{equation}
where $n_c$ is the concentration of identical scatters. Our scattering system is 
sufficiently dilute so that multiple scattering events can be neglected. For
ballistic transport and elastic scattering with system size comparable with 
the mean free path: 
$\mathcal{L}_s\approx\mathcal{L}_{\rm mean-free}=v_F \tau_{tr}$, 
the semiclassical Boltzmann transport theory gives that the conductivity 
is inverse of the $\sigma_{tr}$:
\begin{equation}
    G\propto\frac{1}{\sigma_{tr}}.
\end{equation}
The spin polarization is defined by the spin-resolved transmission coefficient 
as~\cite{TAN2012141}
\begin{align} \nonumber
	P_z=(T^{\downarrow}-T^{\uparrow})/(T^{\downarrow}+T^{\uparrow}). 
\end{align}
We thus have
\begin{equation}
    P_z = \frac{\sigma_{tr}^{\downarrow}-\sigma_{tr}^{\uparrow}}{\sigma_{tr}^{\downarrow}+\sigma_{tr}^{\uparrow}},
\end{equation}
with $\sigma^{\downarrow\uparrow}_{tr}\propto R^{\downarrow\uparrow}$, where the 
resistance R is the inverse of the conductivity $G$.

\section{Validation of S-matrix approach}

\subsection{Reduction from eccentric circular to annular cavity}

For an annular scattering cavity ($\xi=0$), the scattering matrix can be 
analytically calculated, providing a way to validate the scattering-matrix approach
to the general case of $\xi \ne 0$. For this purpose, we consider the annular 
scattering cavity $\xi=0$ but with two boundaries: one at $R_1$ and another at 
$R_2$. In the three regions, the wave functions associated with an angular 
momentum channel are 
\begin{align}\nonumber
    \Psi^{\mathrm{I}}_m&={^{k_{\mathrm{I}}}}h^{(2)}_m+S_{mm}~{^{k_{\mathrm{I}}}}h^{(1)}_m,\\\nonumber
    \Psi^{\mathrm{II}}_m&=A_{m}[{^{k_{\mathrm{II}}}}h^{(2)}_m+S^{cd}_{mm}~{^{k_{\mathrm{II}}}}h^{(1)}_m],\\
    \Psi^{\mathrm{III}}_{m}&=B_m{^{k_{\mathrm{III}}}}j_{m}.
\end{align}
Imposing the boundary conditions at $r=R_1$ and $r = R_2$ gives
\begin{widetext}
    \begin{align}
    \left[\begin{array}{cccc}
    {^{k_{\mathrm{II}}}_{2}}h^{(2)}_m(R_1) &0 &{^{k_{\mathrm{II}}}_{2}}h^{(1)}_m(R_1) &-{^{k_{\mathrm{I}}}_{2}}h^{(1)}_m(R_1)\\
    {^{k_{\mathrm{II}}}_{1+3}}h^{(2)}_m(R_1) &0 &{^{k_{\mathrm{II}}}_{1+3}}h^{(1)}_m(R_1) &-{^{k_{\mathrm{I}}}_{1+3}}h^{(1)}_m(R_1)\\
    {^{k_{\mathrm{II}}}_{2}}h^{(2)}_m(R_2) &-{^{k_{\mathrm{III}}}_{2}}j_m(R_2) &{^{k_{\mathrm{II}}}_{2}}h^{(1)}_m(R_2) &0\\
     {^{k_{\mathrm{II}}}_{1+3}}h^{(2)}_m(R_2) &-{^{k_{\mathrm{III}}}_{1+3}}j_m(R_2) &{^{k_{\mathrm{II}}}_{1+3}}h^{(1)}_m(R_2) &0
    \end{array}
    \right]\left[\begin{array}{c}
        A_{m} \\
        B_{m}\\
        C_{m}\\
        S_{m}
    \end{array}\right]=\left[\begin{array}{c}
        {^{k_{\mathrm{I}}}_{2}}h^{(2)}_m(R_1) \\
        {^{k_{\mathrm{I}}}_{1+3}}h^{(2)}_m(R_1) \\
        0\\
        0
    \end{array}\right],
\end{align}
\end{widetext}
where $C_m\equiv A_m S^{cd}_{mm}$ and
\begin{align}\nonumber
    {^{k_{i}}_2}h^{(1,2)}_{m}(R_j) &= s_i H^{(1,2)}_m(k_{i}R_j), \\ \nonumber
    {^{k_{i}}_{1+3}}h^{(1,2)}_{m}(R_j) &= H^{(1,2)}_{m-1}(k_{i}R_j)-H^{(1,2)}_{m+1}(k_{i}R_j), \\ \nonumber
    {^{k_{i}}_2}j_{m}(R_j) &= s_i J_m(k_{i}R_j), \\ \nonumber
    {^{k_{i}}_{1+3}}j_{m}(R_j) &= J_{m-1}(k_{i}R_j)-J_{m+1}(k_{i}R_j)
\end{align}
with $i=\mathrm{I},\mathrm{II},\mathrm{III}$ and $j=1,2$. Note that 
${^{k_{\mathrm{II}}}_{2}}h^{(2)}_m(R_1)$ and 
${^{k_{\mathrm{II}}}_{1+3}}h^{(2)}_m(R_1)$ are scalars, roughly corresponding to 
the second component and the sum of the first and third components of the radial 
part of ${^{k_{\mathrm{II}}}}h^{(2)}_m(R_1)$, respectively. We have
\begin{align}
    A_m &= \frac{s_{\mathrm{I}}H^{(2)}_m(k_{\mathrm{I}}R_1)+s_{\mathrm{I}}H^{(1)}_m(k_{\mathrm{I}}R_1)S_{mm}}{s_{\mathrm{II}}H^{(2)}_m(k_{\mathrm{II}}R_1)+s_{\mathrm{II}}H^{(1)}_m(k_{\mathrm{II}}R_1)S^{cd}_{mm}}, \\
    B_m&=A_m\frac{s_{\mathrm{II}}H^{(2)}_m(k_{\mathrm{II}}R_2)+s_{\mathrm{II}}H^{(1)}_m(k_{\mathrm{II}}R_2)S^{cd}_{mm}}{s_{\mathrm{III}}J_m(k_{\mathrm{III}}R_2)}.
\end{align}
The scattering matrix is given by
\begin{widetext}
    \begin{equation}\label{eq:SmatrixRing}
    S_{mm}=-\frac{s_{\mathrm{I}}y_m H^{(2)}_m(k_{\mathrm{I}}R_1)-s_{\mathrm{II}}x_m[H^{(2)}_{m-1}(k_{\mathrm{I}}R_1)-H^{(2)}_{m+1}(k_{\mathrm{I}}R_1)]}{s_{\mathrm{I}}y_m H^{(1)}_m(k_{\mathrm{I}}R_1)-s_{\mathrm{II}}x_m[H^{(1)}_{m-1}(k_{\mathrm{I}}R_1)-H^{(1)}_{m+1}(k_{\mathrm{I}}R_1)]},
\end{equation}
where
\begin{align}\nonumber
    x_m &= H^{(2)}_m(k_{\mathrm{II}}R_1)+H^{(1)}_m(k_{\mathrm{II}}R_1)S^{cd}_{mm}\\\nonumber
    y_m &=[H^{(2)}_{m-1}(k_{\mathrm{II}}R_1)-H^{(2)}_{m+1}(k_{\mathrm{II}}R_1)]+[H^{(1)}_{m-1}(k_{\mathrm{II}}R_1)-H^{(1)}_{m+1}(k_{\mathrm{II}}R_1)]S^{cd}_{mm}.
\end{align}
\end{widetext}
For the eccentric circular cavity, the scattering matrix can be determined by 
Eq.~(\ref{eq:SmatrixEccentric}). We can reduce the eccentric cavity to an annular 
cavity by taking the limit $\xi\rightarrow 0$. In that case, the off-diagonal 
scattering matrix will reduce to the diagonal matrix: 
$S^{od}_{lj}\rightarrow S^{cd}_{ll}\delta_{jl}$ and $S_{mm'}\rightarrow S_{mm}\delta_{mm'}$.
We have that Eq.~(\ref{eq:SmatrixEccentric}) reduces to the same form of 
Eq.~(\ref{eq:SmatrixRing}) as
\begin{equation}\label{eq:Sreduce}
    S_{mm}=-\frac{Z^{(2)}_m-Y^{(2)}_m-s_{\mathrm{I}}s_{\mathrm{II}}X^{(2)}_m\mathcal{T}_m}{Z^{(1)}_m-Y^{(1)}_m-s_{\mathrm{I}}s_{\mathrm{II}}X^{(1)}_m\mathcal{T}_m}
\end{equation}
with $\mathcal{T}_m=y_m/x_m$. We find that, numerically, the difference between 
the scattering matrix in Eq.~(\ref{eq:Sreduce}) and that in 
Eq.~(\ref{eq:SmatrixRing}) is on the order of computer round-off error 
(about $10^{-15}$). The excellent agreement between the analytic S-matrix for 
$\xi = 0$ and the numerically calculated matrix in the limit $\xi\rightarrow 0$ 
validates the S-matrix approach manifested through Eq.~(\ref{eq:SmatrixEccentric}).

\begin{figure} [ht!]
\centering
\includegraphics[width=0.8\linewidth]{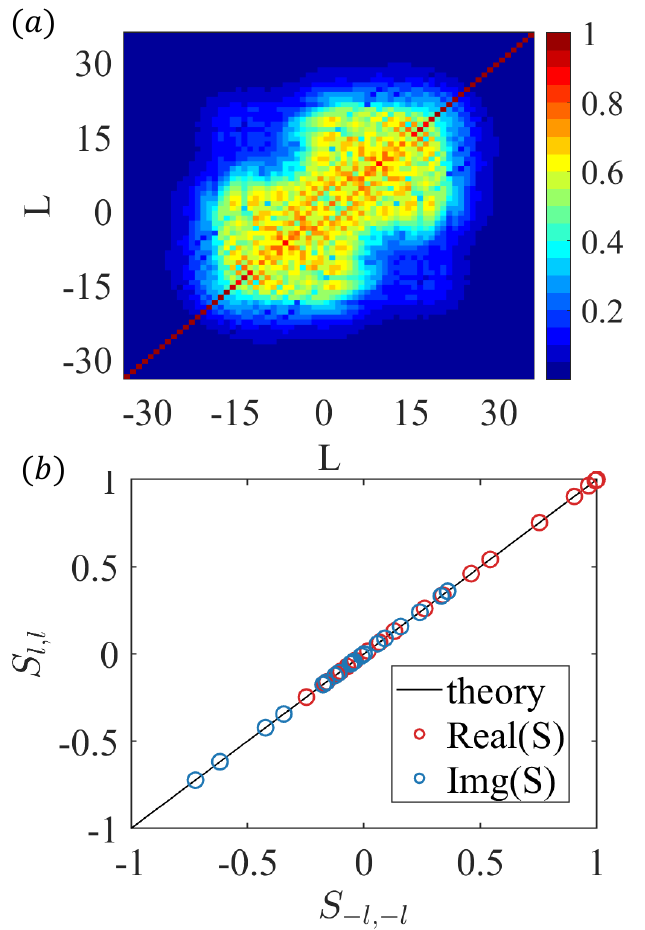}
\caption{Validation of the S-matrix approach. (a) Distribution of the elements
of $|S|^{1/4}$ in a large angular momentum interval. Since the matrix elements 
in the S-matrix are between zero and one, the elements of $|S|^{1/4}$ are used
for better visualization. (b) Mirror symmetry constraint of the real (red) and 
imaginary (blue) parts of the S-matrix elements for $\xi=0.165$. Other parameters 
are: potentials $V_1=-10-2\mu$, $V_2=40-2\mu$ with $\mu=20$ for the spin-up 
electrons. The angular momentum range is $L=-35:1:35$ and the Fermi energy range is 
$E=14.5:(10^{-3}/2):15.3$.}  
\label{fig:validSmatrix}
\end{figure}

\begin{figure*} [ht!]
\centering
\includegraphics[width=0.75\linewidth]{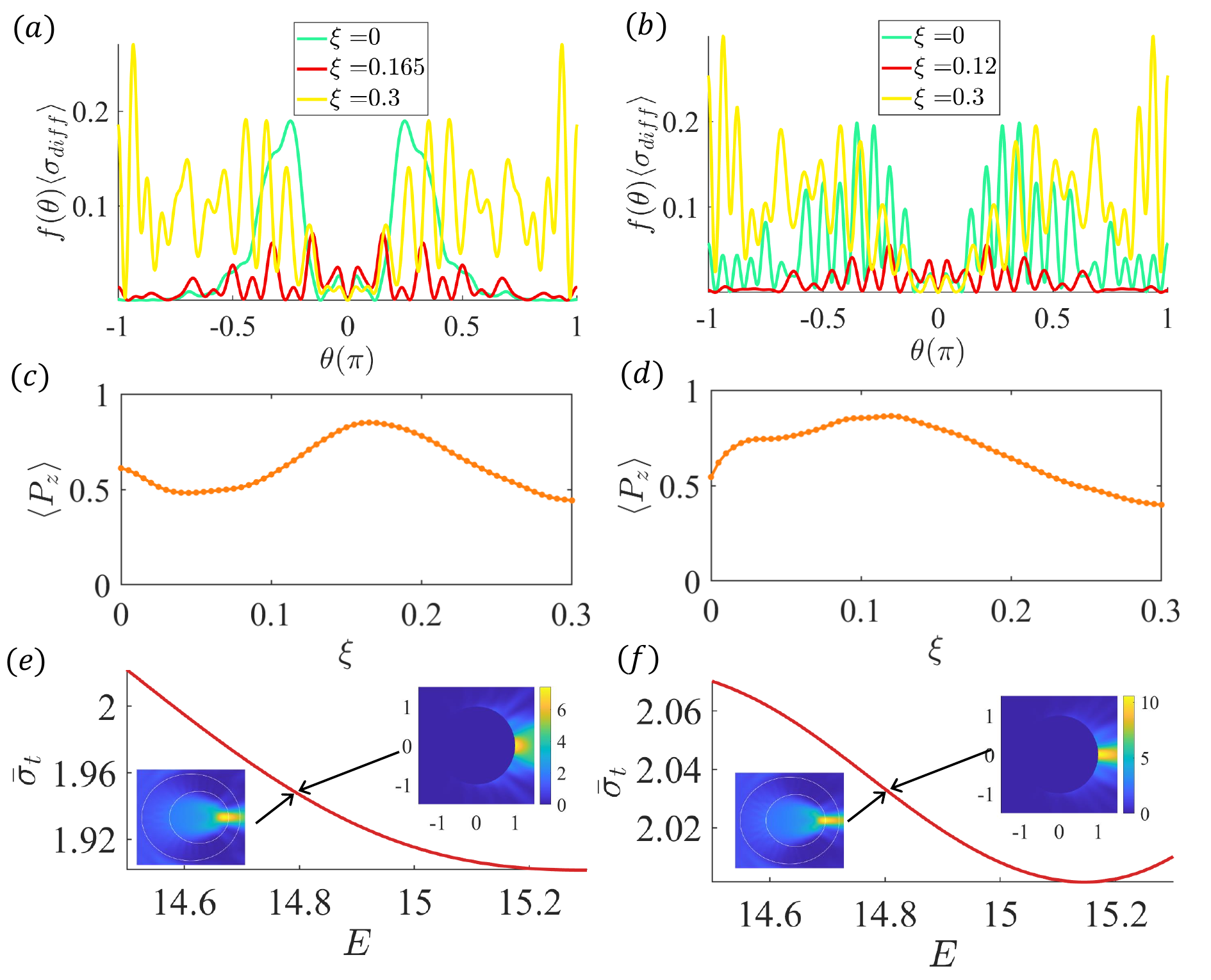}
\caption{Lensing-like pattern for a spin-up Dirac fermion. (a,b) The quantity 
$f(\theta)\langle\sigma_{\rm diff}\rangle$ versus $\theta$ for $\mu=20$ and 
$\mu=22$, respectively, where $f(\theta)\equiv 1-\cos\theta$ and the differential 
cross section $\sigma_{\rm diff}$ is averaged over the Fermi energy. In each case, results from three values 
of $\xi$ are displayed. (c,d) Average spin polarization $\langle P_z\rangle$ over 
Fermi energy versus the eccentric parameter $\xi$ for $\mu=20$ and $\mu=22$, where the 
maximum value of $\langle P_z\rangle$ occurs at $\xi=0.165$ and $0.12$, 
respectively. (e,f) Average total cross section $\bar{\sigma}_t$ versus Fermi energy for 
$\mu=20$ and $\mu=22$, respectively. In each panel, the lower-left inset shows the 
probability distribution while the upper-right inset displays the scattering 
probability distribution for the specific energy value as indicated by the arrows. 
Other parameters are the same as those in Fig.~\ref{fig:EdgeLensing} in the main text. These modes 
have a well-defined classical correspondence: the second kind of classical lensing 
ray pattern satisfying the conditions $\mathcal{C}_2$ (to be detailed in 
Appendix~\ref{sec:ray}).} 
\label{fig:suppFig2}
\end{figure*}

\begin{figure*} [ht!]
\centering
\includegraphics[width=0.7\linewidth]{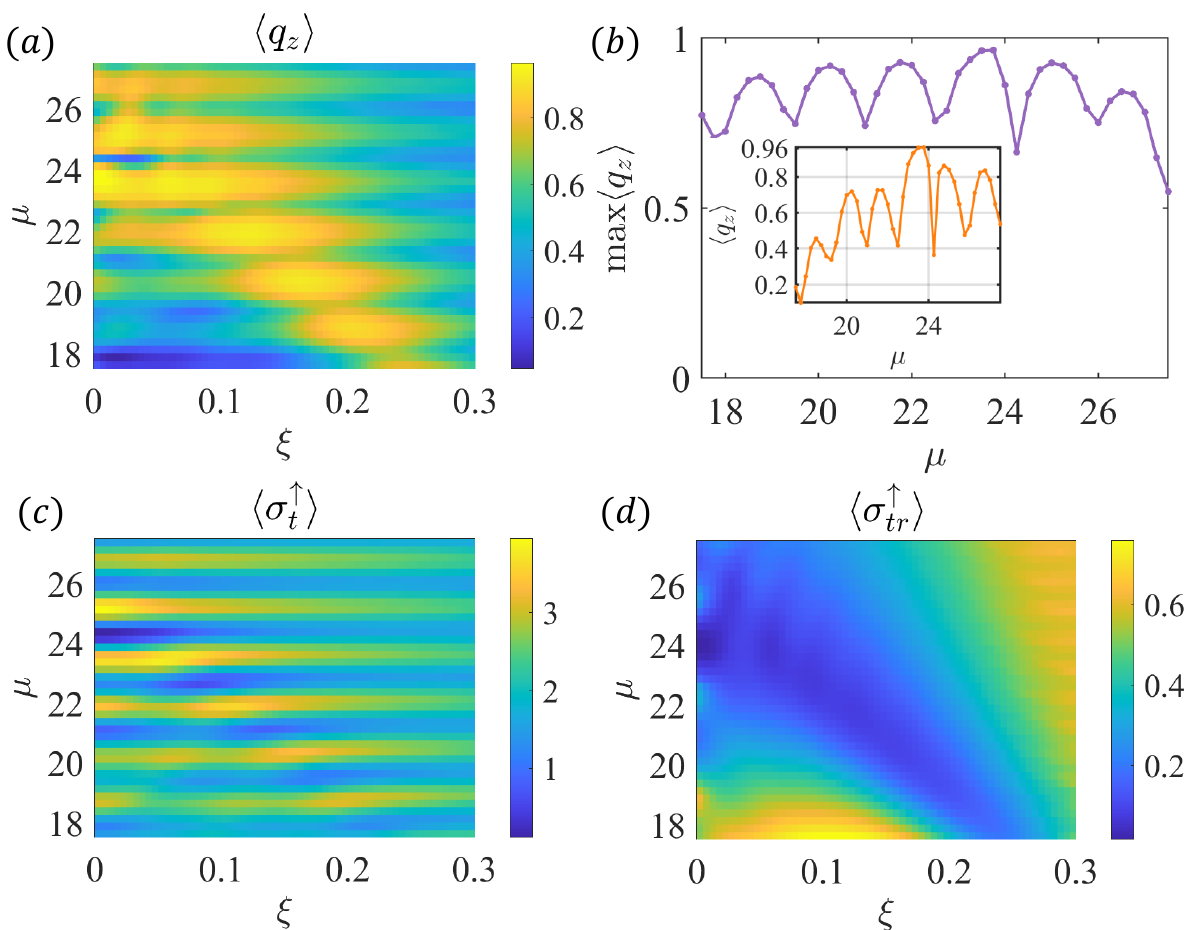}
\caption{Scattering-direction dependent spin polarization. (a) Average 
direction-dependent spin polarization $\langle q_z\rangle$ in the $(\xi,\mu)$
parameter plane. (b) The maximum value $\textnormal{max}\langle q_z\rangle$
(over $\xi$) versus $\mu$. The inset corresponds to the case of $\xi=0$. 
(c) The removed average scattering background cross section 
$\langle \sigma^{\uparrow}_t\rangle$ in the $(\xi,\mu)$ plane. (d) Average 
momentum-transport cross section $\langle \sigma^{\uparrow}_{tr}\rangle$ in the 
$(\xi,\mu)$ plane.}
\label{fig:suppFig3}
\end{figure*}

\begin{figure*} [ht!]
\centering
\includegraphics[width=0.7\linewidth]{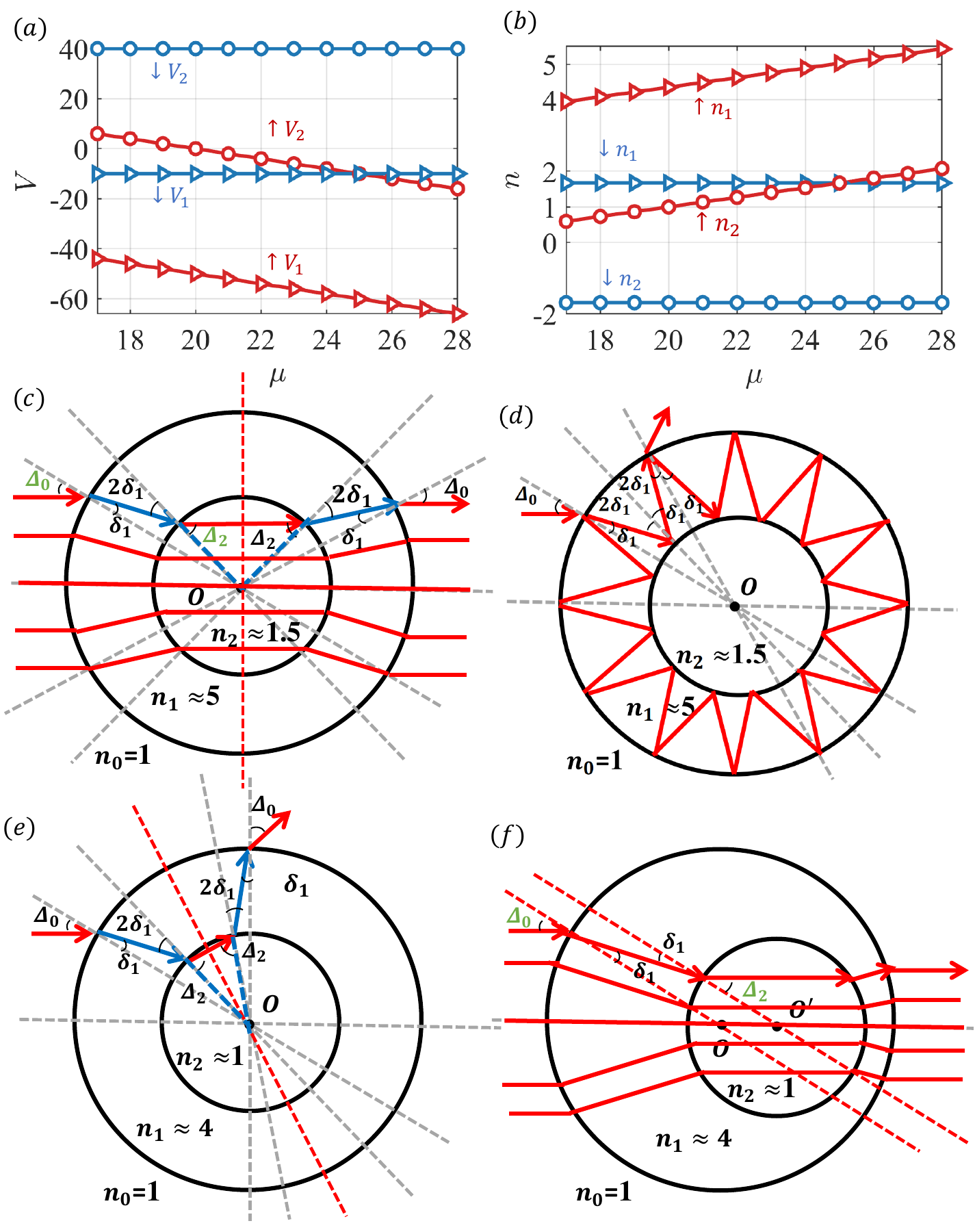}
\caption{Geometric optics interpretation for the quantum lensing-like scattering 
states associated with spin-up Dirac fermions. (a) Total potential versus the exchange potential $\mu$ for spin-down and spin-up
electrons. (b) Effective refractive index in the small 
wavelength limit versus $\mu$ for spin-down and spin-up 
electrons. (c) For $\mu=24$, the annular circular cavity displays one classic 
lensing pattern analogous to the one produced by two convex lenses. An incident 
ray with $\Delta_0<\Delta^c_0$ exhibits no significant scattering.
(d) For $\Delta_0\geq\Delta^c_0$, total internal reflection occurs at the inner 
interface between regions $\mathrm{II}$ and $\mathrm{III}$. However, at the outer 
boundary, total internal reflection does not occur under the condition 
$\mathcal{C}_1$ (see text) and spin-resolved Snell's law, generating a broad 
scattering angle distribution. (e) For $\mu=20$, an annular cavity produces large 
scattering angles even for $\Delta_0<\Delta^c_0$ due to the distinct refractive 
index configuration, away from both conditions $\mathcal{C}_1$ and $\mathcal{C}_2$.
(f) For $\mu=20$, an eccentric circular cavity generates a different kind of 
lensing pattern analogous to the one created by one convex and another concave 
lens.}
\label{fig:rayoptics}
\end{figure*}

Figure~\ref{fig:validSmatrix}(a) shows the convergence of the S-matrix in a large 
angular momentum range. For large angular momenta, the S-matrix elements are 
negligibly small, suggesting that these angular-momentum channels contribute little
to the scattering process. More specifically, Fig.~\ref{fig:validSmatrix}(a) 
is the color map of the S-matrix elements in the angular momentum representation. 
The near-zero components in the high angular momentum basis mean convergence.
Note that the diagonal term in Fig.~\ref{fig:validSmatrix}(a) will be removed in 
the transmission matrix $\mathcal{T}=S-I$, which determines the scattering cross 
sections.

\subsection{Mirror Symmetry}

An eccentric circular cavity possesses the mirror (parity) symmetry. The parity 
operator for pseudospin-$1/2$ quasiparticles is given by~\cite{xu:2018} 
$\mathcal{P}_x=i\sigma_xR_y$, where $R_y$ denotes the mirror transform in the 
position space [e.g., $x\rightarrow x$ ($k_x\rightarrow k_x$),  
$y\rightarrow-y$ ($k_y\rightarrow -k_y$), and $\theta\rightarrow-\theta$], and 
$i\sigma_x$ arises from the $-\pi$ rotation in the counterclockwise direction 
about the $x$ axis in the spin space, i.e., $e^{i\pi\sigma_x/2}$, which is 
equivalent to the mirror-transform operation in the three-dimensional space. For 
pseudospin-$1$ Dirac-Weyl quasiparticles, using Rodrigues’ rotation 
formula~\cite{curtright:2014}, we obtain the rotation operator as
\begin{equation}
e^{i\theta(\hat{n}\cdot\mathbf{J})}=I_3+i(\hat{n}\cdot\mathbf{J})\sin\theta+(\hat{n}\cdot\mathbf{J})^2(\cos\theta-1),
\end{equation}
where $\mathbf{J}$ denotes the total angular momentum, $\hat{n}$ specifies the 
rotation axis and $\theta$ is the rotation angle in the clockwise direction 
around $\hat{n}$. Consider the rotation operation with $\theta=\pi$ around $x$ axis, we have
$\widetilde{S}_x\equiv e^{i\pi S_x} = I_3 -2S^2_x$, so
\begin{align}\nonumber
    \widetilde{S}_x=-\left(\begin{array}{ccc}
    0&0&1\\
    0&1&0\\
    1&0&0
    \end{array}\right),
\end{align}
where $\widetilde{S}_x^2=I_3$ and $\widetilde{S}_x^{-1}=\widetilde{S}_x$.
The parity operator is given by $\mathcal{P}_x = \widetilde{S}_xR_y$ with 
$\mathcal{P}_x\mathcal{P}_x^{-1}=I_3$. The Hamiltonian (\ref{eq:H_s}) is invariant 
under this parity operation:
\begin{align} \nonumber
\mathcal{P}_x\hat{H}_s\mathcal{P}_x^{-1}&=v_F\mathcal{P}_x\mathbf{S}\cdot\mathbf{\hat{p}}\mathcal{P}_x^{-1}+\mathcal{V}_{gate}(x,-y)-s\mathcal{M}(x,-y) \\
&=v_F(S_x\hat{p}_x+S_y\hat{p}_{y})+\mathcal{V}_{gate}(x,-y)-s\mathcal{M}(x,-y).
\end{align}
where 
\begin{align}\nonumber
\mathcal{P}_xS_x\mathcal{P}_x^{-1}\mathcal{P}_x
\hat{p}_x\mathcal{P}_x^{-1}&=S_x\hat{p}_x,\\
\mathcal{P}_xS_y\mathcal{P}_x^{-1}\mathcal{P}_x
\hat{p}_y\mathcal{P}_x^{-1}&=(-S_y)(-\hat{p}_y)=S_y\hat{p}_y.
\end{align}
Finally, using 
\begin{align}\nonumber
    \mathcal{V}_{gate}(x,-y) = \mathcal{V}_{gate}(x,y),\\
    \mathcal{M}(x,-y) = \mathcal{M}(x,y),\nonumber
\end{align}
we obtain $\mathcal{P}_x\hat{H}_s\mathcal{P}_x^{-1}=\hat{H}_s$. As a result, the 
parity operation on the wave function is also a solution of the system. The 
cylindrical spinor basis under the parity transform has the form
\begin{align}
    \mathcal{P}_x~{^{k}}h^{(1,2)}_m&=(-1)^{m+1}\frac{1}{\sqrt{2}}\left(\begin{array}{c}
        H^{(1,2)}_{-m-1}(kr)e^{-i\theta} \\
        is\sqrt{2}H^{(1,2)}_{-m}(kr)\\
        -H^{(1,2)}_{-m+1}(kr)e^{i\theta}
    \end{array}\right)e^{-im\theta}\\\nonumber
    &=(-1)^{m+1}~{^{k}}h^{(1,2)}_{-m}.
\end{align}
The wave function in region $\mathrm{I}$ in the eccentric circular cavity is
\begin{widetext}
    \begin{align}\nonumber
    \mathcal{P}_x\Psi^{\mathrm{I}}&=\sum_{m=-\infty}^{+\infty}\mathcal{P}_xa^{\mathrm{I}}_m\mathcal{P}_x^{-1}\left[\mathcal{P}_x{^{k_{\mathrm{I}}}}h^{(2)}_m+\sum_{m'=-\infty}^{+\infty}\mathcal{P}_xS_{mm'}\mathcal{P}_x^{-1}\mathcal{P}_x{^{k_{\mathrm{I}}}}h^{(1)}_{m'}\right]\\\nonumber
    &=\sum_{m=-\infty}^{+\infty}\mathcal{P}_xa^{\mathrm{I}}_m\mathcal{P}_x^{-1}(-1)^{m+1}\left[{^{k_{\mathrm{I}}}}h^{(2)}_{-m}+\sum_{m'=-\infty}^{+\infty}\mathcal{P}_xS_{mm'}\mathcal{P}_x^{-1}(-1)^{m'-m}~{^{k_{\mathrm{I}}}}h^{(1)}_{-m'}\right]\\
    &=\sum_{n=-\infty}^{+\infty}\widetilde{A}^{\mathrm{I}}_n\left[{^{k_{\mathrm{I}}}}h^{(2)}_{n}+\sum_{n'=-\infty}^{+\infty}S_{nn'}{^{k_{\mathrm{I}}}}h^{(1)}_{n'}\right]
\end{align}
\end{widetext}
with $n\equiv-m$, $n'\equiv-m'$ and $\widetilde{A}^{\mathrm{I}}_n\equiv\mathcal{P}_xa^{\mathrm{I}}_{-n}\mathcal{P}_x^{-1}(-1)^{-n+1}$. We have 
\begin{align}\nonumber
    S_{nn'}&\equiv S_{-m,-m'}\\\nonumber
    &=\mathcal{P}_xS_{mm'}\mathcal{P}_x^{-1}(-1)^{m'-m}\\
    &=(-1)^{m'-m}S_{mm'}.
\end{align}
For $m=m'$, we get $S_{m,m}=S_{-m,-m}$. Thus the real and imaginary parts of 
the S-matrix obey this relation, as shown Fig.~\ref{fig:validSmatrix}(b).

\section{Supplementary information for Fig.~\ref{fig:polar} in the main text}\label{sec:mu_20_22}

For Fig.~\ref{fig:polar}(b) in the main text, the top left inset shows a mode corresponding to 
a classical geometric optic lensing pattern. This type of lensing-like mode 
can shrink an incident parallel beam into a narrow parallel emission flow, as 
shown in Fig.~\ref{fig:suppFig2}.

\section{Scattering-direction dependent spin polarization}\label{sec:detail_polarization}

The average momentum-transport cross section is defined as
\begin{equation}
    \langle \sigma_{tr} \rangle=\frac{1}{E_1-E_0} \int^{E_1}_{E_0} dE~\int^{2\pi}_{0} d \theta ~(1-\cos\theta) |f(\theta,\theta')|^2,
\end{equation}
where $|f(\theta,\theta')|^2$ is the probability for scattering associated with 
incident angle $\theta'$ and scattering angle $\theta$ as in 
Eq.~\eqref{eq:f_theta}. The weighting factor $1-\cos\theta$ is used to 
quantify the scattering angle deviation from the incident angle $\theta'=0$. The 
quantity $\langle \sigma_{tr} \rangle$ contains two implicit parts: the total 
scattering probability (or scattering background $\sigma_t$) and the scattering 
angle $\theta$ distribution with respect to the incident direction. To separate the 
effect of scattering direction on spin polarization, we remove the background by 
normalizing the scattering probability over $\theta$ with the total scattering 
cross section, $|f(\theta,\theta')|^2/\sigma_t$, and define an alternative spin 
polarization that depends on the scattering direction as
\begin{equation}
    q_z = \frac{\sigma^{\downarrow}_{tr}/\sigma^{\downarrow}_t-\sigma^{\uparrow}_{tr}/\sigma^{\uparrow}_t}{\sigma^{\downarrow}_{tr}/\sigma^{\downarrow}_t+\sigma^{\uparrow}_{tr}/\sigma^{\uparrow}_t},
\end{equation}
where the ratio $\sigma_{tr}/\sigma_t$ is proportional to the average momentum 
transfer cross section~\cite{enwiki:1116845797} over the scattering angle with
\begin{equation}
\langle\Delta\mathbf{p}\rangle_{\Omega}=q\hat{\mathbf{x}}\sigma_{tr}/\sigma_{t},
\end{equation}
where the incident direction is along $\hat{\mathbf{x}}$, 
$\Delta\mathbf{p}\equiv\mathbf{p}_{in}-\mathbf{p}_{out}$, $q$ is the incident momentum 
magnitude, and $\Omega$ is the scattering solid angle. Figure~\ref{fig:suppFig3}(a) 
shows the numerically calculated $\langle q_z\rangle$ over Fermi energy in the parameter 
plane $(\xi,\mu)$. It can be seen that high spin polarization can be achieved. 
Figure~\ref{fig:suppFig3}(b) shows, the maximum value 
$\textnormal{max}\langle q_z\rangle$ versus $\mu$, which can be as large as $96\%$! 
Figure~\ref{fig:suppFig3}(c) shows the removed average scattering background cross 
section $\langle \sigma^{\uparrow}_t\rangle$ in the $(\xi,\mu)$ plane, which 
exhibits a periodic structure in $\mu$. For reference, Fig.~\ref{fig:suppFig3}(d) 
shows the average momentum-transport cross section 
$\langle \sigma^{\uparrow}_{tr}\rangle$ in the $(\xi,\mu)$ plane.

\section{Understanding spin-up fermion lensing modes based on Dirac electron optics} \label{sec:ray}

We provide a geometric-optics-based interpretation to understand the lensing-like 
scattering states associated with spin-up Dirac fermions through two kinds of 
classical lensing ray patterns. Figures~\ref{fig:rayoptics}(a) and 
\ref{fig:rayoptics}(b) show the total potential and effective refractive index
versus the exchange potential $\mu$, respectively, for spin-down and spin-up
electrons. 

The set of conditions under which the first type of classical lensing ray pattern 
arises (denoted as $\mathcal{C}_1$), as shown in Fig.~\ref{fig:rayoptics}(c), is: 
(1) an infinitesimal refractive angle $\delta_1$, (2) approximately equal lengths 
of the solid and dashed blue ray paths, (3) $\Delta_2\approx\Delta_0+\delta\Delta$ 
with infinitesimal term $\delta\Delta$, and 
(4) $\Delta_0<\Delta^c_0 <\pi/2$. The Snell's law, 
$\sin\Delta_2\approx 2\delta_1 n_{1}/n_2$
and $\sin\Delta_0\approx \delta_1 n_{1}/n_0$, gives 
\begin{align} \nonumber
\sin(\Delta_0+\delta\Delta)\approx\sin(\Delta_0) 2n_{0}/n_{2}\gtrsim \sin(\Delta_0). 
\end{align}
Condition $\mathcal{C}_1$ requires $2n_0/n_2\gtrsim 1$, so the refractive index 
$n_2$ should be at least $n_2\approx 2-\delta n\lesssim2$. The first kind of 
classical lensing pattern displayed in Fig.~\ref{fig:rayoptics}(c) corresponds to 
the $\mu=24$ case with the effective refractive index in the small wavelength 
limit: $n_2\lesssim2$, $n_1\approx5$, and $n_0=1$, as shown in 
Fig.~\ref{fig:rayoptics}(b). In this case, the average spin polarization 
$\langle P_z\rangle$ reaches maximum for the annular cavity with $\xi=0$ and the 
ray pattern in Fig.~\ref{fig:rayoptics}(c) resembles the scattering probability 
of the corresponding lensing-like mode in the left panel in Fig.~2(d) in the main 
text for $E=14.8$. For $\mu=24$, the critic incident angle is determined by 
\begin{align} \nonumber
\sin\Delta^c_0\approx \delta^c_1~n_1/n_0\approx 0.75, 
\end{align}
with $\sin\Delta_2=2\delta^c_1~n_1/n_2=1$, so $\Delta^c_0\approx48.6^o$. For 
$\Delta_0\geq\Delta^c_0$, total internal reflections occur at the inner interface 
between regions $\mathrm{II}$ and $\mathrm{III}$ but will not at the outer 
boundary, resulting in a vast scattering angle distribution, as shown in 
Fig.~\ref{fig:rayoptics}(d), which resembles the pattern with the resonant 
quantum state in the Fermi energy range of lensing-like modes in Fig.~2(b) in the main 
text. In principle, the directional distribution of the leaking of the quantum 
resonant states can be quantitatively understood by semi-classical 
simulation~\cite{wiersig:2008,schrepfer:2021}. 

The set of conditions $\mathcal{C}_2$ under which the second type of lensing 
pattern arises, as shown in Fig.~\ref{fig:rayoptics}(f), is: (1) infinitesimal 
refractive angle $\delta_1$, (2) the two red dashed ray segments in 
Fig.~\ref{fig:rayoptics}(f) being approximately parallel, 
(3) $\Delta_2\approx\Delta_0$, and (4) $\Delta_0<\Delta^c_0 <\pi/2$. 
Starting from the conditions $\mathcal{C}_1$, if $n_2$ is away from two, such as 
for $\mu=20$ with $n_2\approx 1$, the annular cavity shape produces large 
scattering angles because of the large deviation of $\Delta_2$ from $\Delta_0$, 
as shown in Fig.~\ref{fig:rayoptics}(e), breaking both conditions $\mathcal{C}_1$ 
and $\mathcal{C}_2$. For the potential configuration with $\mu=20$, condition 
$\mathcal{C}_2$ is satisfied for an eccentric circular cavity, producing the 
lensing pattern in Fig.~\ref{fig:rayoptics}(f), which resembles the corresponding 
lensing-like mode in the insets of Fig.~\ref{fig:suppFig2}(e,f) and the upper 
inset of Fig.~3(b) in the main text. The corresponding critical incident angle is 
$\Delta^c_0\approx 30^o$, which is smaller than that in the $\mu=24$ case.

In general, total internal reflections disrupt parallel rays, where a small 
critical incident angle will generate a large spread of the emitted rays. While 
all rays in the effective refractive index configuration associated with the 
classical-quantum correspondence for $\mu\in[20,24]$ can produce the classical 
lensing ray pattern with the proper incident angle and eccentric parameter $\xi$, 
an enlarged critical angle is indicative of the contribution to scattering from
the lensing patterns. As a result, in the corresponding quantum regime, the spin 
polarization increases from $\mu=20$ to $\mu=24$. In principle, if $\mu$ is 
increased further, the corresponding classical lensing ray pattern will occur 
for $\xi<0$ and generate patterns similar to those for $\xi>0$. 

We note that the edge states of spin-down electrons break the ray-wave 
correspondence~\cite{xu:2019}, their scattering behaviors cannot be explained 
by geometric optics.

\bibliography{spinPolar}

\begin{thebibliography}{68}%
\makeatletter
\providecommand \@ifxundefined [1]{%
 \@ifx{#1\undefined}
}%
\providecommand \@ifnum [1]{%
 \ifnum #1\expandafter \@firstoftwo
 \else \expandafter \@secondoftwo
 \fi
}%
\providecommand \@ifx [1]{%
 \ifx #1\expandafter \@firstoftwo
 \else \expandafter \@secondoftwo
 \fi
}%
\providecommand \natexlab [1]{#1}%
\providecommand \enquote  [1]{``#1''}%
\providecommand \bibnamefont  [1]{#1}%
\providecommand \bibfnamefont [1]{#1}%
\providecommand \citenamefont [1]{#1}%
\providecommand \href@noop [0]{\@secondoftwo}%
\providecommand \href [0]{\begingroup \@sanitize@url \@href}%
\providecommand \@href[1]{\@@startlink{#1}\@@href}%
\providecommand \@@href[1]{\endgroup#1\@@endlink}%
\providecommand \@sanitize@url [0]{\catcode `\\12\catcode `\$12\catcode `\&12\catcode `\#12\catcode `\^12\catcode `\_12\catcode `\%12\relax}%
\providecommand \@@startlink[1]{}%
\providecommand \@@endlink[0]{}%
\providecommand \url  [0]{\begingroup\@sanitize@url \@url }%
\providecommand \@url [1]{\endgroup\@href {#1}{\urlprefix }}%
\providecommand \urlprefix  [0]{URL }%
\providecommand \Eprint [0]{\href }%
\providecommand \doibase [0]{https://doi.org/}%
\providecommand \selectlanguage [0]{\@gobble}%
\providecommand \bibinfo  [0]{\@secondoftwo}%
\providecommand \bibfield  [0]{\@secondoftwo}%
\providecommand \translation [1]{[#1]}%
\providecommand \BibitemOpen [0]{}%
\providecommand \bibitemStop [0]{}%
\providecommand \bibitemNoStop [0]{.\EOS\space}%
\providecommand \EOS [0]{\spacefactor3000\relax}%
\providecommand \BibitemShut  [1]{\csname bibitem#1\endcsname}%
\let\auto@bib@innerbib\@empty
%</preamble>
\bibitem [{\citenamefont {Betancur-Ocampo}\ \emph {et~al.}(2019)\citenamefont {Betancur-Ocampo}, \citenamefont {Leyvraz},\ and\ \citenamefont {Stegmann}}]{betancur:2019}%
  \BibitemOpen
  \bibfield  {author} {\bibinfo {author} {\bibfnamefont {Y.}~\bibnamefont {Betancur-Ocampo}}, \bibinfo {author} {\bibfnamefont {F.}~\bibnamefont {Leyvraz}},\ and\ \bibinfo {author} {\bibfnamefont {T.}~\bibnamefont {Stegmann}},\ }\bibfield  {title} {\bibinfo {title} {Electron optics in phosphorene pn junctions: negative reflection and anti-{super-Klein} tunneling},\ }\href@noop {} {\bibfield  {journal} {\bibinfo  {journal} {Nano. Lett.}\ }\textbf {\bibinfo {volume} {19}},\ \bibinfo {pages} {7760} (\bibinfo {year} {2019})}\BibitemShut {NoStop}%
\bibitem [{\citenamefont {Xu}\ \emph {et~al.}(2018)\citenamefont {Xu}, \citenamefont {Wang}, \citenamefont {Huang},\ and\ \citenamefont {Lai}}]{xu:2018}%
  \BibitemOpen
  \bibfield  {author} {\bibinfo {author} {\bibfnamefont {H.-Y.}\ \bibnamefont {Xu}}, \bibinfo {author} {\bibfnamefont {G.-L.}\ \bibnamefont {Wang}}, \bibinfo {author} {\bibfnamefont {L.}~\bibnamefont {Huang}},\ and\ \bibinfo {author} {\bibfnamefont {Y.-C.}\ \bibnamefont {Lai}},\ }\bibfield  {title} {\bibinfo {title} {Chaos in {Dirac} electron optics: Emergence of a relativistic {Quantum Chimera}},\ }\href {https://doi.org/10.1103/PhysRevLett.120.124101} {\bibfield  {journal} {\bibinfo  {journal} {Phys. Rev. Lett.}\ }\textbf {\bibinfo {volume} {120}},\ \bibinfo {pages} {124101} (\bibinfo {year} {2018})}\BibitemShut {NoStop}%
\bibitem [{\citenamefont {Cheianov}\ \emph {et~al.}(2007)\citenamefont {Cheianov}, \citenamefont {Fal'ko},\ and\ \citenamefont {Altshuler}}]{cheianov:2007}%
  \BibitemOpen
  \bibfield  {author} {\bibinfo {author} {\bibfnamefont {V.~V.}\ \bibnamefont {Cheianov}}, \bibinfo {author} {\bibfnamefont {V.}~\bibnamefont {Fal'ko}},\ and\ \bibinfo {author} {\bibfnamefont {B.}~\bibnamefont {Altshuler}},\ }\bibfield  {title} {\bibinfo {title} {The focusing of electron flow and a {Veselago} lens in graphene pn junctions},\ }\href@noop {} {\bibfield  {journal} {\bibinfo  {journal} {Science}\ }\textbf {\bibinfo {volume} {315}},\ \bibinfo {pages} {1252} (\bibinfo {year} {2007})}\BibitemShut {NoStop}%
\bibitem [{\citenamefont {Cserti}\ \emph {et~al.}(2007)\citenamefont {Cserti}, \citenamefont {P{\'a}lyi},\ and\ \citenamefont {P{\'e}terfalvi}}]{cserti:2007}%
  \BibitemOpen
  \bibfield  {author} {\bibinfo {author} {\bibfnamefont {J.}~\bibnamefont {Cserti}}, \bibinfo {author} {\bibfnamefont {A.}~\bibnamefont {P{\'a}lyi}},\ and\ \bibinfo {author} {\bibfnamefont {C.}~\bibnamefont {P{\'e}terfalvi}},\ }\bibfield  {title} {\bibinfo {title} {Caustics due to a negative refractive index in circular graphene p- n junctions},\ }\href@noop {} {\bibfield  {journal} {\bibinfo  {journal} {Phys. Rev. Lett.}\ }\textbf {\bibinfo {volume} {99}},\ \bibinfo {pages} {246801} (\bibinfo {year} {2007})}\BibitemShut {NoStop}%
\bibitem [{\citenamefont {Wang}\ \emph {et~al.}(2019{\natexlab{a}})\citenamefont {Wang}, \citenamefont {Elahi}, \citenamefont {Wang}, \citenamefont {Habib}, \citenamefont {Taniguchi}, \citenamefont {Watanabe}, \citenamefont {Hone}, \citenamefont {Ghosh}, \citenamefont {Lee},\ and\ \citenamefont {Kim}}]{wang:2019}%
  \BibitemOpen
  \bibfield  {author} {\bibinfo {author} {\bibfnamefont {K.}~\bibnamefont {Wang}}, \bibinfo {author} {\bibfnamefont {M.~M.}\ \bibnamefont {Elahi}}, \bibinfo {author} {\bibfnamefont {L.}~\bibnamefont {Wang}}, \bibinfo {author} {\bibfnamefont {K.~M.}\ \bibnamefont {Habib}}, \bibinfo {author} {\bibfnamefont {T.}~\bibnamefont {Taniguchi}}, \bibinfo {author} {\bibfnamefont {K.}~\bibnamefont {Watanabe}}, \bibinfo {author} {\bibfnamefont {J.}~\bibnamefont {Hone}}, \bibinfo {author} {\bibfnamefont {A.~W.}\ \bibnamefont {Ghosh}}, \bibinfo {author} {\bibfnamefont {G.-H.}\ \bibnamefont {Lee}},\ and\ \bibinfo {author} {\bibfnamefont {P.}~\bibnamefont {Kim}},\ }\bibfield  {title} {\bibinfo {title} {Graphene transistor based on tunable {Dirac} fermion optics},\ }\href@noop {} {\bibfield  {journal} {\bibinfo  {journal} {Proc. Natl. Acad. Sci. (USA)}\ }\textbf {\bibinfo {volume} {116}},\ \bibinfo {pages} {6575} (\bibinfo {year} {2019}{\natexlab{a}})}\BibitemShut {NoStop}%
\bibitem [{\citenamefont {Shytov}\ \emph {et~al.}(2008)\citenamefont {Shytov}, \citenamefont {Rudner},\ and\ \citenamefont {Levitov}}]{shytov:2008}%
  \BibitemOpen
  \bibfield  {author} {\bibinfo {author} {\bibfnamefont {A.~V.}\ \bibnamefont {Shytov}}, \bibinfo {author} {\bibfnamefont {M.~S.}\ \bibnamefont {Rudner}},\ and\ \bibinfo {author} {\bibfnamefont {L.~S.}\ \bibnamefont {Levitov}},\ }\bibfield  {title} {\bibinfo {title} {Klein backscattering and {Fabry-P{\'e}rot} interference in graphene heterojunctions},\ }\href@noop {} {\bibfield  {journal} {\bibinfo  {journal} {Phys. Rev. Lett.}\ }\textbf {\bibinfo {volume} {101}},\ \bibinfo {pages} {156804} (\bibinfo {year} {2008})}\BibitemShut {NoStop}%
\bibitem [{\citenamefont {Rickhaus}\ \emph {et~al.}(2013)\citenamefont {Rickhaus}, \citenamefont {Maurand}, \citenamefont {Liu}, \citenamefont {Weiss}, \citenamefont {Richter},\ and\ \citenamefont {Sch{\"o}nenberger}}]{rickhaus:2013}%
  \BibitemOpen
  \bibfield  {author} {\bibinfo {author} {\bibfnamefont {P.}~\bibnamefont {Rickhaus}}, \bibinfo {author} {\bibfnamefont {R.}~\bibnamefont {Maurand}}, \bibinfo {author} {\bibfnamefont {M.-H.}\ \bibnamefont {Liu}}, \bibinfo {author} {\bibfnamefont {M.}~\bibnamefont {Weiss}}, \bibinfo {author} {\bibfnamefont {K.}~\bibnamefont {Richter}},\ and\ \bibinfo {author} {\bibfnamefont {C.}~\bibnamefont {Sch{\"o}nenberger}},\ }\bibfield  {title} {\bibinfo {title} {Ballistic interferences in suspended graphene},\ }\href@noop {} {\bibfield  {journal} {\bibinfo  {journal} {Nat. Commun.}\ }\textbf {\bibinfo {volume} {4}},\ \bibinfo {pages} {2342} (\bibinfo {year} {2013})}\BibitemShut {NoStop}%
\bibitem [{\citenamefont {Gu}\ \emph {et~al.}(2011)\citenamefont {Gu}, \citenamefont {Rudner},\ and\ \citenamefont {Levitov}}]{gu:2011}%
  \BibitemOpen
  \bibfield  {author} {\bibinfo {author} {\bibfnamefont {N.}~\bibnamefont {Gu}}, \bibinfo {author} {\bibfnamefont {M.}~\bibnamefont {Rudner}},\ and\ \bibinfo {author} {\bibfnamefont {L.}~\bibnamefont {Levitov}},\ }\bibfield  {title} {\bibinfo {title} {Chirality-assisted electronic cloaking of confined states in bilayer graphene},\ }\href@noop {} {\bibfield  {journal} {\bibinfo  {journal} {Phys. Rev. Lett.}\ }\textbf {\bibinfo {volume} {107}},\ \bibinfo {pages} {156603} (\bibinfo {year} {2011})}\BibitemShut {NoStop}%
\bibitem [{\citenamefont {B{\o}ggild}\ \emph {et~al.}(2017)\citenamefont {B{\o}ggild}, \citenamefont {Caridad}, \citenamefont {Stampfer}, \citenamefont {Calogero}, \citenamefont {Papior},\ and\ \citenamefont {Brandbyge}}]{boggild:2017}%
  \BibitemOpen
  \bibfield  {author} {\bibinfo {author} {\bibfnamefont {P.}~\bibnamefont {B{\o}ggild}}, \bibinfo {author} {\bibfnamefont {J.~M.}\ \bibnamefont {Caridad}}, \bibinfo {author} {\bibfnamefont {C.}~\bibnamefont {Stampfer}}, \bibinfo {author} {\bibfnamefont {G.}~\bibnamefont {Calogero}}, \bibinfo {author} {\bibfnamefont {N.~R.}\ \bibnamefont {Papior}},\ and\ \bibinfo {author} {\bibfnamefont {M.}~\bibnamefont {Brandbyge}},\ }\bibfield  {title} {\bibinfo {title} {A two-dimensional {Dirac} fermion microscope},\ }\href@noop {} {\bibfield  {journal} {\bibinfo  {journal} {Nat. Commun.}\ }\textbf {\bibinfo {volume} {8}},\ \bibinfo {pages} {15783} (\bibinfo {year} {2017})}\BibitemShut {NoStop}%
\bibitem [{\citenamefont {Heinisch}\ \emph {et~al.}(2013)\citenamefont {Heinisch}, \citenamefont {Bronold},\ and\ \citenamefont {Fehske}}]{heinisch:2013}%
  \BibitemOpen
  \bibfield  {author} {\bibinfo {author} {\bibfnamefont {R.}~\bibnamefont {Heinisch}}, \bibinfo {author} {\bibfnamefont {F.}~\bibnamefont {Bronold}},\ and\ \bibinfo {author} {\bibfnamefont {H.}~\bibnamefont {Fehske}},\ }\bibfield  {title} {\bibinfo {title} {Mie scattering analog in graphene: {Lensing}, particle confinement, and depletion of {Klein} tunneling},\ }\href@noop {} {\bibfield  {journal} {\bibinfo  {journal} {Phys. Rev. B}\ }\textbf {\bibinfo {volume} {87}},\ \bibinfo {pages} {155409} (\bibinfo {year} {2013})}\BibitemShut {NoStop}%
\bibitem [{\citenamefont {Caridad}\ \emph {et~al.}(2016)\citenamefont {Caridad}, \citenamefont {Connaughton}, \citenamefont {Ott}, \citenamefont {Weber},\ and\ \citenamefont {Krsti{\'c}}}]{caridad:2016}%
  \BibitemOpen
  \bibfield  {author} {\bibinfo {author} {\bibfnamefont {J.~M.}\ \bibnamefont {Caridad}}, \bibinfo {author} {\bibfnamefont {S.}~\bibnamefont {Connaughton}}, \bibinfo {author} {\bibfnamefont {C.}~\bibnamefont {Ott}}, \bibinfo {author} {\bibfnamefont {H.~B.}\ \bibnamefont {Weber}},\ and\ \bibinfo {author} {\bibfnamefont {V.}~\bibnamefont {Krsti{\'c}}},\ }\bibfield  {title} {\bibinfo {title} {An electrical analogy to {Mie} scattering},\ }\href@noop {} {\bibfield  {journal} {\bibinfo  {journal} {Nat. Commun.}\ }\textbf {\bibinfo {volume} {7}},\ \bibinfo {pages} {12894} (\bibinfo {year} {2016})}\BibitemShut {NoStop}%
\bibitem [{\citenamefont {Guti{\'e}rrez}\ \emph {et~al.}(2016)\citenamefont {Guti{\'e}rrez}, \citenamefont {Brown}, \citenamefont {Kim}, \citenamefont {Park},\ and\ \citenamefont {Pasupathy}}]{gutierrez:2016}%
  \BibitemOpen
  \bibfield  {author} {\bibinfo {author} {\bibfnamefont {C.}~\bibnamefont {Guti{\'e}rrez}}, \bibinfo {author} {\bibfnamefont {L.}~\bibnamefont {Brown}}, \bibinfo {author} {\bibfnamefont {C.-J.}\ \bibnamefont {Kim}}, \bibinfo {author} {\bibfnamefont {J.}~\bibnamefont {Park}},\ and\ \bibinfo {author} {\bibfnamefont {A.~N.}\ \bibnamefont {Pasupathy}},\ }\bibfield  {title} {\bibinfo {title} {Klein tunnelling and electron trapping in nanometre-scale graphene quantum dots},\ }\href@noop {} {\bibfield  {journal} {\bibinfo  {journal} {Nat. Phys.}\ }\textbf {\bibinfo {volume} {12}},\ \bibinfo {pages} {1069} (\bibinfo {year} {2016})}\BibitemShut {NoStop}%
\bibitem [{\citenamefont {Lee}\ \emph {et~al.}(2016)\citenamefont {Lee}, \citenamefont {Wong}, \citenamefont {Velasco~Jr}, \citenamefont {Rodriguez-Nieva}, \citenamefont {Kahn}, \citenamefont {Tsai}, \citenamefont {Taniguchi}, \citenamefont {Watanabe}, \citenamefont {Zettl}, \citenamefont {Wang} \emph {et~al.}}]{lee:2016}%
  \BibitemOpen
  \bibfield  {author} {\bibinfo {author} {\bibfnamefont {J.}~\bibnamefont {Lee}}, \bibinfo {author} {\bibfnamefont {D.}~\bibnamefont {Wong}}, \bibinfo {author} {\bibfnamefont {J.}~\bibnamefont {Velasco~Jr}}, \bibinfo {author} {\bibfnamefont {J.~F.}\ \bibnamefont {Rodriguez-Nieva}}, \bibinfo {author} {\bibfnamefont {S.}~\bibnamefont {Kahn}}, \bibinfo {author} {\bibfnamefont {H.-Z.}\ \bibnamefont {Tsai}}, \bibinfo {author} {\bibfnamefont {T.}~\bibnamefont {Taniguchi}}, \bibinfo {author} {\bibfnamefont {K.}~\bibnamefont {Watanabe}}, \bibinfo {author} {\bibfnamefont {A.}~\bibnamefont {Zettl}}, \bibinfo {author} {\bibfnamefont {F.}~\bibnamefont {Wang}}, \emph {et~al.},\ }\bibfield  {title} {\bibinfo {title} {Imaging electrostatically confined {Dirac} fermions in graphene quantum dots},\ }\href@noop {} {\bibfield  {journal} {\bibinfo  {journal} {Nat. Phys.}\ }\textbf {\bibinfo {volume} {12}},\ \bibinfo {pages} {1032} (\bibinfo {year} {2016})}\BibitemShut {NoStop}%
\bibitem [{\citenamefont {Sadrara}\ and\ \citenamefont {Miri}(2019)}]{sadrara:2019}%
  \BibitemOpen
  \bibfield  {author} {\bibinfo {author} {\bibfnamefont {M.}~\bibnamefont {Sadrara}}\ and\ \bibinfo {author} {\bibfnamefont {M.}~\bibnamefont {Miri}},\ }\bibfield  {title} {\bibinfo {title} {Dirac electron scattering from a cluster of electrostatically defined quantum dots in graphene},\ }\href@noop {} {\bibfield  {journal} {\bibinfo  {journal} {Phys. Rev. B}\ }\textbf {\bibinfo {volume} {99}},\ \bibinfo {pages} {155432} (\bibinfo {year} {2019})}\BibitemShut {NoStop}%
\bibitem [{\citenamefont {Nguyen}\ and\ \citenamefont {Charlier}(2018)}]{nguyen:2018}%
  \BibitemOpen
  \bibfield  {author} {\bibinfo {author} {\bibfnamefont {V.~H.}\ \bibnamefont {Nguyen}}\ and\ \bibinfo {author} {\bibfnamefont {J.-C.}\ \bibnamefont {Charlier}},\ }\bibfield  {title} {\bibinfo {title} {Klein tunneling and electron optics in {Dirac-Weyl} fermion systems with tilted energy dispersion},\ }\href@noop {} {\bibfield  {journal} {\bibinfo  {journal} {Phys. Rev. B}\ }\textbf {\bibinfo {volume} {97}},\ \bibinfo {pages} {235113} (\bibinfo {year} {2018})}\BibitemShut {NoStop}%
\bibitem [{\citenamefont {Reijnders}\ \emph {et~al.}(2018)\citenamefont {Reijnders}, \citenamefont {Minenkov}, \citenamefont {Katsnelson},\ and\ \citenamefont {Dobrokhotov}}]{reijnders:2018}%
  \BibitemOpen
  \bibfield  {author} {\bibinfo {author} {\bibfnamefont {K.}~\bibnamefont {Reijnders}}, \bibinfo {author} {\bibfnamefont {D.}~\bibnamefont {Minenkov}}, \bibinfo {author} {\bibfnamefont {M.}~\bibnamefont {Katsnelson}},\ and\ \bibinfo {author} {\bibfnamefont {S.~Y.}\ \bibnamefont {Dobrokhotov}},\ }\bibfield  {title} {\bibinfo {title} {Electronic optics in graphene in the semiclassical approximation},\ }\href@noop {} {\bibfield  {journal} {\bibinfo  {journal} {Ann. Phys.}\ }\textbf {\bibinfo {volume} {397}},\ \bibinfo {pages} {65} (\bibinfo {year} {2018})}\BibitemShut {NoStop}%
\bibitem [{\citenamefont {Brun}\ \emph {et~al.}(2019)\citenamefont {Brun}, \citenamefont {Moreau}, \citenamefont {Somanchi}, \citenamefont {Nguyen}, \citenamefont {Watanabe}, \citenamefont {Taniguchi}, \citenamefont {Charlier}, \citenamefont {Stampfer},\ and\ \citenamefont {Hackens}}]{brun:2019}%
  \BibitemOpen
  \bibfield  {author} {\bibinfo {author} {\bibfnamefont {B.}~\bibnamefont {Brun}}, \bibinfo {author} {\bibfnamefont {N.}~\bibnamefont {Moreau}}, \bibinfo {author} {\bibfnamefont {S.}~\bibnamefont {Somanchi}}, \bibinfo {author} {\bibfnamefont {V.-H.}\ \bibnamefont {Nguyen}}, \bibinfo {author} {\bibfnamefont {K.}~\bibnamefont {Watanabe}}, \bibinfo {author} {\bibfnamefont {T.}~\bibnamefont {Taniguchi}}, \bibinfo {author} {\bibfnamefont {J.-C.}\ \bibnamefont {Charlier}}, \bibinfo {author} {\bibfnamefont {C.}~\bibnamefont {Stampfer}},\ and\ \bibinfo {author} {\bibfnamefont {B.}~\bibnamefont {Hackens}},\ }\bibfield  {title} {\bibinfo {title} {Imaging {Dirac} fermions flow through a circular {Veselago} lens},\ }\href@noop {} {\bibfield  {journal} {\bibinfo  {journal} {Phys. Rev. B}\ }\textbf {\bibinfo {volume} {100}},\ \bibinfo {pages} {041401} (\bibinfo {year} {2019})}\BibitemShut {NoStop}%
\bibitem [{\citenamefont {Bai}\ \emph {et~al.}(2018)\citenamefont {Bai}, \citenamefont {Zhou}, \citenamefont {Wei}, \citenamefont {Qiao}, \citenamefont {Liu}, \citenamefont {Liu}, \citenamefont {Jiang},\ and\ \citenamefont {He}}]{bai:2018}%
  \BibitemOpen
  \bibfield  {author} {\bibinfo {author} {\bibfnamefont {K.-K.}\ \bibnamefont {Bai}}, \bibinfo {author} {\bibfnamefont {J.-J.}\ \bibnamefont {Zhou}}, \bibinfo {author} {\bibfnamefont {Y.-C.}\ \bibnamefont {Wei}}, \bibinfo {author} {\bibfnamefont {J.-B.}\ \bibnamefont {Qiao}}, \bibinfo {author} {\bibfnamefont {Y.-W.}\ \bibnamefont {Liu}}, \bibinfo {author} {\bibfnamefont {H.-W.}\ \bibnamefont {Liu}}, \bibinfo {author} {\bibfnamefont {H.}~\bibnamefont {Jiang}},\ and\ \bibinfo {author} {\bibfnamefont {L.}~\bibnamefont {He}},\ }\bibfield  {title} {\bibinfo {title} {Generating atomically sharp p- n junctions in graphene and testing quantum electron optics on the nanoscale},\ }\href@noop {} {\bibfield  {journal} {\bibinfo  {journal} {Phys. Rev. B}\ }\textbf {\bibinfo {volume} {97}},\ \bibinfo {pages} {045413} (\bibinfo {year} {2018})}\BibitemShut {NoStop}%
\bibitem [{\citenamefont {Wei}\ \emph {et~al.}(2016)\citenamefont {Wei}, \citenamefont {Lee}, \citenamefont {Lemaitre}, \citenamefont {Pinel}, \citenamefont {Cutaia}, \citenamefont {Cha}, \citenamefont {Katmis}, \citenamefont {Zhu}, \citenamefont {Heiman}, \citenamefont {Hone} \emph {et~al.}}]{wei:2016}%
  \BibitemOpen
  \bibfield  {author} {\bibinfo {author} {\bibfnamefont {P.}~\bibnamefont {Wei}}, \bibinfo {author} {\bibfnamefont {S.}~\bibnamefont {Lee}}, \bibinfo {author} {\bibfnamefont {F.}~\bibnamefont {Lemaitre}}, \bibinfo {author} {\bibfnamefont {L.}~\bibnamefont {Pinel}}, \bibinfo {author} {\bibfnamefont {D.}~\bibnamefont {Cutaia}}, \bibinfo {author} {\bibfnamefont {W.}~\bibnamefont {Cha}}, \bibinfo {author} {\bibfnamefont {F.}~\bibnamefont {Katmis}}, \bibinfo {author} {\bibfnamefont {Y.}~\bibnamefont {Zhu}}, \bibinfo {author} {\bibfnamefont {D.}~\bibnamefont {Heiman}}, \bibinfo {author} {\bibfnamefont {J.}~\bibnamefont {Hone}}, \emph {et~al.},\ }\bibfield  {title} {\bibinfo {title} {Strong interfacial exchange field in the {graphene/EuS} heterostructure},\ }\href@noop {} {\bibfield  {journal} {\bibinfo  {journal} {Nat. Mater.}\ }\textbf {\bibinfo {volume} {15}},\ \bibinfo {pages} {711} (\bibinfo {year} {2016})}\BibitemShut {NoStop}%
\bibitem [{\citenamefont {Singh}\ \emph {et~al.}(2017)\citenamefont {Singh}, \citenamefont {Katoch}, \citenamefont {Zhu}, \citenamefont {Meng}, \citenamefont {Liu}, \citenamefont {Brangham}, \citenamefont {Yang}, \citenamefont {Flatt{\'e}},\ and\ \citenamefont {Kawakami}}]{singh:2017}%
  \BibitemOpen
  \bibfield  {author} {\bibinfo {author} {\bibfnamefont {S.}~\bibnamefont {Singh}}, \bibinfo {author} {\bibfnamefont {J.}~\bibnamefont {Katoch}}, \bibinfo {author} {\bibfnamefont {T.}~\bibnamefont {Zhu}}, \bibinfo {author} {\bibfnamefont {K.-Y.}\ \bibnamefont {Meng}}, \bibinfo {author} {\bibfnamefont {T.}~\bibnamefont {Liu}}, \bibinfo {author} {\bibfnamefont {J.~T.}\ \bibnamefont {Brangham}}, \bibinfo {author} {\bibfnamefont {F.}~\bibnamefont {Yang}}, \bibinfo {author} {\bibfnamefont {M.~E.}\ \bibnamefont {Flatt{\'e}}},\ and\ \bibinfo {author} {\bibfnamefont {R.~K.}\ \bibnamefont {Kawakami}},\ }\bibfield  {title} {\bibinfo {title} {Strong modulation of spin currents in bilayer graphene by static and fluctuating proximity exchange fields},\ }\href@noop {} {\bibfield  {journal} {\bibinfo  {journal} {Phys. Rev. Lett.}\ }\textbf {\bibinfo {volume} {118}},\ \bibinfo {pages} {187201} (\bibinfo {year} {2017})}\BibitemShut {NoStop}%
\bibitem [{\citenamefont {Haugen}\ \emph {et~al.}(2008)\citenamefont {Haugen}, \citenamefont {Huertas-Hernando},\ and\ \citenamefont {Brataas}}]{haugen:2008}%
  \BibitemOpen
  \bibfield  {author} {\bibinfo {author} {\bibfnamefont {H.}~\bibnamefont {Haugen}}, \bibinfo {author} {\bibfnamefont {D.}~\bibnamefont {Huertas-Hernando}},\ and\ \bibinfo {author} {\bibfnamefont {A.}~\bibnamefont {Brataas}},\ }\bibfield  {title} {\bibinfo {title} {Spin transport in proximity-induced ferromagnetic graphene},\ }\href@noop {} {\bibfield  {journal} {\bibinfo  {journal} {Phys. Rev. B}\ }\textbf {\bibinfo {volume} {77}},\ \bibinfo {pages} {115406} (\bibinfo {year} {2008})}\BibitemShut {NoStop}%
\bibitem [{\citenamefont {Yang}\ \emph {et~al.}(2013)\citenamefont {Yang}, \citenamefont {Hallal}, \citenamefont {Terrade}, \citenamefont {Waintal}, \citenamefont {Roche},\ and\ \citenamefont {Chshiev}}]{yang:2013}%
  \BibitemOpen
  \bibfield  {author} {\bibinfo {author} {\bibfnamefont {H.-X.}\ \bibnamefont {Yang}}, \bibinfo {author} {\bibfnamefont {A.}~\bibnamefont {Hallal}}, \bibinfo {author} {\bibfnamefont {D.}~\bibnamefont {Terrade}}, \bibinfo {author} {\bibfnamefont {X.}~\bibnamefont {Waintal}}, \bibinfo {author} {\bibfnamefont {S.}~\bibnamefont {Roche}},\ and\ \bibinfo {author} {\bibfnamefont {M.}~\bibnamefont {Chshiev}},\ }\bibfield  {title} {\bibinfo {title} {Proximity effects induced in graphene by magnetic insulators: first-principles calculations on spin filtering and exchange-splitting gaps},\ }\href@noop {} {\bibfield  {journal} {\bibinfo  {journal} {Phys. Rev. Lett.}\ }\textbf {\bibinfo {volume} {110}},\ \bibinfo {pages} {046603} (\bibinfo {year} {2013})}\BibitemShut {NoStop}%
\bibitem [{\citenamefont {Li}\ \emph {et~al.}(2013)\citenamefont {Li}, \citenamefont {Roschewsky}, \citenamefont {Assaf}, \citenamefont {Eich}, \citenamefont {Epstein-Martin}, \citenamefont {Heiman}, \citenamefont {M{\"u}nzenberg},\ and\ \citenamefont {Moodera}}]{li:2013}%
  \BibitemOpen
  \bibfield  {author} {\bibinfo {author} {\bibfnamefont {B.}~\bibnamefont {Li}}, \bibinfo {author} {\bibfnamefont {N.}~\bibnamefont {Roschewsky}}, \bibinfo {author} {\bibfnamefont {B.~A.}\ \bibnamefont {Assaf}}, \bibinfo {author} {\bibfnamefont {M.}~\bibnamefont {Eich}}, \bibinfo {author} {\bibfnamefont {M.}~\bibnamefont {Epstein-Martin}}, \bibinfo {author} {\bibfnamefont {D.}~\bibnamefont {Heiman}}, \bibinfo {author} {\bibfnamefont {M.}~\bibnamefont {M{\"u}nzenberg}},\ and\ \bibinfo {author} {\bibfnamefont {J.~S.}\ \bibnamefont {Moodera}},\ }\bibfield  {title} {\bibinfo {title} {Superconducting spin switch with infinite magnetoresistance induced by an internal exchange field},\ }\href@noop {} {\bibfield  {journal} {\bibinfo  {journal} {Phys. Rev. Lett.}\ }\textbf {\bibinfo {volume} {110}},\ \bibinfo {pages} {097001} (\bibinfo {year} {2013})}\BibitemShut {NoStop}%
\bibitem [{\citenamefont {Moghaddam}\ and\ \citenamefont {Zareyan}(2010)}]{moghaddam:2010}%
  \BibitemOpen
  \bibfield  {author} {\bibinfo {author} {\bibfnamefont {A.~G.}\ \bibnamefont {Moghaddam}}\ and\ \bibinfo {author} {\bibfnamefont {M.}~\bibnamefont {Zareyan}},\ }\bibfield  {title} {\bibinfo {title} {Graphene-based electronic spin lenses},\ }\href@noop {} {\bibfield  {journal} {\bibinfo  {journal} {Phys. Rev. Lett.}\ }\textbf {\bibinfo {volume} {105}},\ \bibinfo {pages} {146803} (\bibinfo {year} {2010})}\BibitemShut {NoStop}%
\bibitem [{\citenamefont {Grivet}\ \emph {et~al.}(2013)\citenamefont {Grivet}, \citenamefont {Hawkes},\ and\ \citenamefont {Septier}}]{grivet:2013}%
  \BibitemOpen
  \bibfield  {author} {\bibinfo {author} {\bibfnamefont {P.}~\bibnamefont {Grivet}}, \bibinfo {author} {\bibfnamefont {P.~W.}\ \bibnamefont {Hawkes}},\ and\ \bibinfo {author} {\bibfnamefont {A.}~\bibnamefont {Septier}},\ }\href@noop {} {\emph {\bibinfo {title} {Electron optics}}}\ (\bibinfo  {publisher} {Elsevier},\ \bibinfo {year} {2013})\BibitemShut {NoStop}%
\bibitem [{\citenamefont {Batson}\ \emph {et~al.}(2002)\citenamefont {Batson}, \citenamefont {Dellby},\ and\ \citenamefont {Krivanek}}]{batson:2002}%
  \BibitemOpen
  \bibfield  {author} {\bibinfo {author} {\bibfnamefont {P.~E.}\ \bibnamefont {Batson}}, \bibinfo {author} {\bibfnamefont {N.}~\bibnamefont {Dellby}},\ and\ \bibinfo {author} {\bibfnamefont {O.~L.}\ \bibnamefont {Krivanek}},\ }\bibfield  {title} {\bibinfo {title} {Sub-{\aa}ngstrom resolution using aberration corrected electron optics},\ }\href@noop {} {\bibfield  {journal} {\bibinfo  {journal} {Nature}\ }\textbf {\bibinfo {volume} {418}},\ \bibinfo {pages} {617} (\bibinfo {year} {2002})}\BibitemShut {NoStop}%
\bibitem [{\citenamefont {Chen}\ \emph {et~al.}(2016)\citenamefont {Chen}, \citenamefont {Han}, \citenamefont {Elahi}, \citenamefont {Habib}, \citenamefont {Wang}, \citenamefont {Wen}, \citenamefont {Gao}, \citenamefont {Taniguchi}, \citenamefont {Watanabe}, \citenamefont {Hone} \emph {et~al.}}]{chen:2016}%
  \BibitemOpen
  \bibfield  {author} {\bibinfo {author} {\bibfnamefont {S.}~\bibnamefont {Chen}}, \bibinfo {author} {\bibfnamefont {Z.}~\bibnamefont {Han}}, \bibinfo {author} {\bibfnamefont {M.~M.}\ \bibnamefont {Elahi}}, \bibinfo {author} {\bibfnamefont {K.~M.}\ \bibnamefont {Habib}}, \bibinfo {author} {\bibfnamefont {L.}~\bibnamefont {Wang}}, \bibinfo {author} {\bibfnamefont {B.}~\bibnamefont {Wen}}, \bibinfo {author} {\bibfnamefont {Y.}~\bibnamefont {Gao}}, \bibinfo {author} {\bibfnamefont {T.}~\bibnamefont {Taniguchi}}, \bibinfo {author} {\bibfnamefont {K.}~\bibnamefont {Watanabe}}, \bibinfo {author} {\bibfnamefont {J.}~\bibnamefont {Hone}}, \emph {et~al.},\ }\bibfield  {title} {\bibinfo {title} {Electron optics with pn junctions in ballistic graphene},\ }\href@noop {} {\bibfield  {journal} {\bibinfo  {journal} {Science}\ }\textbf {\bibinfo {volume} {353}},\ \bibinfo {pages} {1522} (\bibinfo {year} {2016})}\BibitemShut {NoStop}%
\bibitem [{\citenamefont {Tian}\ \emph {et~al.}(2012)\citenamefont {Tian}, \citenamefont {Chan},\ and\ \citenamefont {Wang}}]{tian:2012}%
  \BibitemOpen
  \bibfield  {author} {\bibinfo {author} {\bibfnamefont {H.}~\bibnamefont {Tian}}, \bibinfo {author} {\bibfnamefont {K.~S.}\ \bibnamefont {Chan}},\ and\ \bibinfo {author} {\bibfnamefont {J.}~\bibnamefont {Wang}},\ }\bibfield  {title} {\bibinfo {title} {Efficient spin injection in graphene using electron optics},\ }\href@noop {} {\bibfield  {journal} {\bibinfo  {journal} {Phys. Rev. B}\ }\textbf {\bibinfo {volume} {86}},\ \bibinfo {pages} {245413} (\bibinfo {year} {2012})}\BibitemShut {NoStop}%
\bibitem [{\citenamefont {Wang}\ \emph {et~al.}(2019{\natexlab{b}})\citenamefont {Wang}, \citenamefont {Han}, \citenamefont {Xu},\ and\ \citenamefont {Lai}}]{wang:2019chaos}%
  \BibitemOpen
  \bibfield  {author} {\bibinfo {author} {\bibfnamefont {C.-Z.}\ \bibnamefont {Wang}}, \bibinfo {author} {\bibfnamefont {C.-D.}\ \bibnamefont {Han}}, \bibinfo {author} {\bibfnamefont {H.-Y.}\ \bibnamefont {Xu}},\ and\ \bibinfo {author} {\bibfnamefont {Y.-C.}\ \bibnamefont {Lai}},\ }\bibfield  {title} {\bibinfo {title} {Chaos-based berry phase detector},\ }\href@noop {} {\bibfield  {journal} {\bibinfo  {journal} {Phys. Rev. B}\ }\textbf {\bibinfo {volume} {99}},\ \bibinfo {pages} {144302} (\bibinfo {year} {2019}{\natexlab{b}})}\BibitemShut {NoStop}%
\bibitem [{\citenamefont {Schrepfer}\ \emph {et~al.}(2021)\citenamefont {Schrepfer}, \citenamefont {Chen}, \citenamefont {Liu}, \citenamefont {Richter},\ and\ \citenamefont {Hentschel}}]{schrepfer:2021}%
  \BibitemOpen
  \bibfield  {author} {\bibinfo {author} {\bibfnamefont {J.-K.}\ \bibnamefont {Schrepfer}}, \bibinfo {author} {\bibfnamefont {S.-C.}\ \bibnamefont {Chen}}, \bibinfo {author} {\bibfnamefont {M.-H.}\ \bibnamefont {Liu}}, \bibinfo {author} {\bibfnamefont {K.}~\bibnamefont {Richter}},\ and\ \bibinfo {author} {\bibfnamefont {M.}~\bibnamefont {Hentschel}},\ }\bibfield  {title} {\bibinfo {title} {Dirac fermion optics and directed emission from single- and bilayer graphene cavities},\ }\href {https://doi.org/10.1103/PhysRevB.104.155436} {\bibfield  {journal} {\bibinfo  {journal} {Phys. Rev. B}\ }\textbf {\bibinfo {volume} {104}},\ \bibinfo {pages} {155436} (\bibinfo {year} {2021})}\BibitemShut {NoStop}%
\bibitem [{\citenamefont {Wang}\ and\ \citenamefont {Liu}(2022)}]{wang:2022}%
  \BibitemOpen
  \bibfield  {author} {\bibinfo {author} {\bibfnamefont {J.}~\bibnamefont {Wang}}\ and\ \bibinfo {author} {\bibfnamefont {J.-F.}\ \bibnamefont {Liu}},\ }\bibfield  {title} {\bibinfo {title} {Super-klein tunneling and electron-beam collimation in the honeycomb superlattice},\ }\href@noop {} {\bibfield  {journal} {\bibinfo  {journal} {Phys. Rev. B}\ }\textbf {\bibinfo {volume} {105}},\ \bibinfo {pages} {035402} (\bibinfo {year} {2022})}\BibitemShut {NoStop}%
\bibitem [{\citenamefont {Xu}\ and\ \citenamefont {Lai}(2019)}]{xu:2019}%
  \BibitemOpen
  \bibfield  {author} {\bibinfo {author} {\bibfnamefont {H.-Y.}\ \bibnamefont {Xu}}\ and\ \bibinfo {author} {\bibfnamefont {Y.-C.}\ \bibnamefont {Lai}},\ }\bibfield  {title} {\bibinfo {title} {Pseudospin-1 wave scattering that defies chaos {$Q$}-spoiling and {Klein} tunneling},\ }\href {https://doi.org/10.1103/PhysRevB.99.235403} {\bibfield  {journal} {\bibinfo  {journal} {Phys. Rev. B}\ }\textbf {\bibinfo {volume} {99}},\ \bibinfo {pages} {235403} (\bibinfo {year} {2019})}\BibitemShut {NoStop}%
\bibitem [{\citenamefont {Xu}\ \emph {et~al.}(2021)\citenamefont {Xu}, \citenamefont {Huang},\ and\ \citenamefont {Lai}}]{XHL:2021}%
  \BibitemOpen
  \bibfield  {author} {\bibinfo {author} {\bibfnamefont {H.-Y.}\ \bibnamefont {Xu}}, \bibinfo {author} {\bibfnamefont {L.}~\bibnamefont {Huang}},\ and\ \bibinfo {author} {\bibfnamefont {Y.-C.}\ \bibnamefont {Lai}},\ }\bibfield  {title} {\bibinfo {title} {Klein scattering of spin-1 {Dirac-Weyl} wave and localized surface plasmon},\ }\href {https://doi.org/10.1103/PhysRevResearch.3.013284} {\bibfield  {journal} {\bibinfo  {journal} {Phys. Rev. Res.}\ }\textbf {\bibinfo {volume} {3}},\ \bibinfo {pages} {013284} (\bibinfo {year} {2021})}\BibitemShut {NoStop}%
\bibitem [{\citenamefont {Xu}\ and\ \citenamefont {Lai}(2020{\natexlab{a}})}]{XL:2020a}%
  \BibitemOpen
  \bibfield  {author} {\bibinfo {author} {\bibfnamefont {H.-Y.}\ \bibnamefont {Xu}}\ and\ \bibinfo {author} {\bibfnamefont {Y.-C.}\ \bibnamefont {Lai}},\ }\bibfield  {title} {\bibinfo {title} {Anomalous chiral edge states in spin-1 {Dirac} quantum dots},\ }\href {https://doi.org/10.1103/PhysRevResearch.2.013062} {\bibfield  {journal} {\bibinfo  {journal} {Phys. Rev. Res.}\ }\textbf {\bibinfo {volume} {2}},\ \bibinfo {pages} {013062} (\bibinfo {year} {2020}{\natexlab{a}})}\BibitemShut {NoStop}%
\bibitem [{\citenamefont {Xu}\ and\ \citenamefont {Lai}(2020{\natexlab{b}})}]{XL:2020b}%
  \BibitemOpen
  \bibfield  {author} {\bibinfo {author} {\bibfnamefont {H.-Y.}\ \bibnamefont {Xu}}\ and\ \bibinfo {author} {\bibfnamefont {Y.-C.}\ \bibnamefont {Lai}},\ }\bibfield  {title} {\bibinfo {title} {Anomalous in-gap edge states in two-dimensional pseudospin-1 {Dirac} insulators},\ }\href {https://doi.org/10.1103/PhysRevResearch.2.023368} {\bibfield  {journal} {\bibinfo  {journal} {Phys. Rev. Res.}\ }\textbf {\bibinfo {volume} {2}},\ \bibinfo {pages} {023368} (\bibinfo {year} {2020}{\natexlab{b}})}\BibitemShut {NoStop}%
\bibitem [{\citenamefont {Fang}\ \emph {et~al.}(2016)\citenamefont {Fang}, \citenamefont {Zhang}, \citenamefont {Louie},\ and\ \citenamefont {Chan}}]{fang:2016}%
  \BibitemOpen
  \bibfield  {author} {\bibinfo {author} {\bibfnamefont {A.}~\bibnamefont {Fang}}, \bibinfo {author} {\bibfnamefont {Z.}~\bibnamefont {Zhang}}, \bibinfo {author} {\bibfnamefont {S.~G.}\ \bibnamefont {Louie}},\ and\ \bibinfo {author} {\bibfnamefont {C.~T.}\ \bibnamefont {Chan}},\ }\bibfield  {title} {\bibinfo {title} {Klein tunneling and supercollimation of pseudospin-1 electromagnetic waves},\ }\href@noop {} {\bibfield  {journal} {\bibinfo  {journal} {Phys. Rev. B}\ }\textbf {\bibinfo {volume} {93}},\ \bibinfo {pages} {035422} (\bibinfo {year} {2016})}\BibitemShut {NoStop}%
\bibitem [{\citenamefont {Mukherjee}\ \emph {et~al.}(2015)\citenamefont {Mukherjee}, \citenamefont {Spracklen}, \citenamefont {Choudhury}, \citenamefont {Goldman}, \citenamefont {{\"O}hberg}, \citenamefont {Andersson},\ and\ \citenamefont {Thomson}}]{mukherjee:2015}%
  \BibitemOpen
  \bibfield  {author} {\bibinfo {author} {\bibfnamefont {S.}~\bibnamefont {Mukherjee}}, \bibinfo {author} {\bibfnamefont {A.}~\bibnamefont {Spracklen}}, \bibinfo {author} {\bibfnamefont {D.}~\bibnamefont {Choudhury}}, \bibinfo {author} {\bibfnamefont {N.}~\bibnamefont {Goldman}}, \bibinfo {author} {\bibfnamefont {P.}~\bibnamefont {{\"O}hberg}}, \bibinfo {author} {\bibfnamefont {E.}~\bibnamefont {Andersson}},\ and\ \bibinfo {author} {\bibfnamefont {R.~R.}\ \bibnamefont {Thomson}},\ }\bibfield  {title} {\bibinfo {title} {Observation of a localized flat-band state in a photonic {Lieb} lattice},\ }\href@noop {} {\bibfield  {journal} {\bibinfo  {journal} {Phys. Rev. Lett.}\ }\textbf {\bibinfo {volume} {114}},\ \bibinfo {pages} {245504} (\bibinfo {year} {2015})}\BibitemShut {NoStop}%
\bibitem [{\citenamefont {Vicencio}\ \emph {et~al.}(2015)\citenamefont {Vicencio}, \citenamefont {Cantillano}, \citenamefont {Morales-Inostroza}, \citenamefont {Real}, \citenamefont {Mej{\'\i}a-Cort{\'e}s}, \citenamefont {Weimann}, \citenamefont {Szameit},\ and\ \citenamefont {Molina}}]{vicencio:2015}%
  \BibitemOpen
  \bibfield  {author} {\bibinfo {author} {\bibfnamefont {R.~A.}\ \bibnamefont {Vicencio}}, \bibinfo {author} {\bibfnamefont {C.}~\bibnamefont {Cantillano}}, \bibinfo {author} {\bibfnamefont {L.}~\bibnamefont {Morales-Inostroza}}, \bibinfo {author} {\bibfnamefont {B.}~\bibnamefont {Real}}, \bibinfo {author} {\bibfnamefont {C.}~\bibnamefont {Mej{\'\i}a-Cort{\'e}s}}, \bibinfo {author} {\bibfnamefont {S.}~\bibnamefont {Weimann}}, \bibinfo {author} {\bibfnamefont {A.}~\bibnamefont {Szameit}},\ and\ \bibinfo {author} {\bibfnamefont {M.~I.}\ \bibnamefont {Molina}},\ }\bibfield  {title} {\bibinfo {title} {Observation of localized states in {Lieb} photonic lattices},\ }\href@noop {} {\bibfield  {journal} {\bibinfo  {journal} {Phys. Rev. Lett.}\ }\textbf {\bibinfo {volume} {114}},\ \bibinfo {pages} {245503} (\bibinfo {year} {2015})}\BibitemShut {NoStop}%
\bibitem [{\citenamefont {Diebel}\ \emph {et~al.}(2016)\citenamefont {Diebel}, \citenamefont {Leykam}, \citenamefont {Kroesen}, \citenamefont {Denz},\ and\ \citenamefont {Desyatnikov}}]{diebel:2016}%
  \BibitemOpen
  \bibfield  {author} {\bibinfo {author} {\bibfnamefont {F.}~\bibnamefont {Diebel}}, \bibinfo {author} {\bibfnamefont {D.}~\bibnamefont {Leykam}}, \bibinfo {author} {\bibfnamefont {S.}~\bibnamefont {Kroesen}}, \bibinfo {author} {\bibfnamefont {C.}~\bibnamefont {Denz}},\ and\ \bibinfo {author} {\bibfnamefont {A.~S.}\ \bibnamefont {Desyatnikov}},\ }\bibfield  {title} {\bibinfo {title} {Conical diffraction and composite {Lieb} bosons in photonic lattices},\ }\href@noop {} {\bibfield  {journal} {\bibinfo  {journal} {Phys. Rev. Lett.}\ }\textbf {\bibinfo {volume} {116}},\ \bibinfo {pages} {183902} (\bibinfo {year} {2016})}\BibitemShut {NoStop}%
\bibitem [{\citenamefont {Malcolm}\ and\ \citenamefont {Nicol}(2016)}]{malcolm:2016}%
  \BibitemOpen
  \bibfield  {author} {\bibinfo {author} {\bibfnamefont {J.}~\bibnamefont {Malcolm}}\ and\ \bibinfo {author} {\bibfnamefont {E.}~\bibnamefont {Nicol}},\ }\bibfield  {title} {\bibinfo {title} {Frequency-dependent polarizability, plasmons, and screening in the two-dimensional pseudospin-1 dice lattice},\ }\href@noop {} {\bibfield  {journal} {\bibinfo  {journal} {Phys. Rev. B}\ }\textbf {\bibinfo {volume} {93}},\ \bibinfo {pages} {165433} (\bibinfo {year} {2016})}\BibitemShut {NoStop}%
\bibitem [{\citenamefont {Tomadin}\ \emph {et~al.}(2018)\citenamefont {Tomadin}, \citenamefont {Hornett}, \citenamefont {Wang}, \citenamefont {Alexeev}, \citenamefont {Candini}, \citenamefont {Coletti}, \citenamefont {Turchinovich}, \citenamefont {Kl{\"a}ui}, \citenamefont {Bonn}, \citenamefont {Koppens} \emph {et~al.}}]{tomadin:2018}%
  \BibitemOpen
  \bibfield  {author} {\bibinfo {author} {\bibfnamefont {A.}~\bibnamefont {Tomadin}}, \bibinfo {author} {\bibfnamefont {S.~M.}\ \bibnamefont {Hornett}}, \bibinfo {author} {\bibfnamefont {H.~I.}\ \bibnamefont {Wang}}, \bibinfo {author} {\bibfnamefont {E.~M.}\ \bibnamefont {Alexeev}}, \bibinfo {author} {\bibfnamefont {A.}~\bibnamefont {Candini}}, \bibinfo {author} {\bibfnamefont {C.}~\bibnamefont {Coletti}}, \bibinfo {author} {\bibfnamefont {D.}~\bibnamefont {Turchinovich}}, \bibinfo {author} {\bibfnamefont {M.}~\bibnamefont {Kl{\"a}ui}}, \bibinfo {author} {\bibfnamefont {M.}~\bibnamefont {Bonn}}, \bibinfo {author} {\bibfnamefont {F.~H.}\ \bibnamefont {Koppens}}, \emph {et~al.},\ }\bibfield  {title} {\bibinfo {title} {The ultrafast dynamics and conductivity of photoexcited graphene at different fermi energies},\ }\href@noop {} {\bibfield  {journal} {\bibinfo  {journal} {Sci. Adv.}\ }\textbf {\bibinfo {volume} {4}},\ \bibinfo {pages} {eaar5313} (\bibinfo {year} {2018})}\BibitemShut {NoStop}%
\bibitem [{\citenamefont {Ullal}\ \emph {et~al.}(2019)\citenamefont {Ullal}, \citenamefont {Shi},\ and\ \citenamefont {Sundararaman}}]{ullal:2019}%
  \BibitemOpen
  \bibfield  {author} {\bibinfo {author} {\bibfnamefont {C.~K.}\ \bibnamefont {Ullal}}, \bibinfo {author} {\bibfnamefont {J.}~\bibnamefont {Shi}},\ and\ \bibinfo {author} {\bibfnamefont {R.}~\bibnamefont {Sundararaman}},\ }\bibfield  {title} {\bibinfo {title} {Electron mobility in graphene without invoking the dirac equation},\ }\href@noop {} {\bibfield  {journal} {\bibinfo  {journal} {Am. J. Phys.}\ }\textbf {\bibinfo {volume} {87}},\ \bibinfo {pages} {291} (\bibinfo {year} {2019})}\BibitemShut {NoStop}%
\bibitem [{\citenamefont {Balgley}\ \emph {et~al.}(2022)\citenamefont {Balgley}, \citenamefont {Butler}, \citenamefont {Biswas}, \citenamefont {Ge}, \citenamefont {Lagasse}, \citenamefont {Taniguchi}, \citenamefont {Watanabe}, \citenamefont {Cothrine}, \citenamefont {Mandrus}, \citenamefont {Velasco~Jr} \emph {et~al.}}]{balgley:2022}%
  \BibitemOpen
  \bibfield  {author} {\bibinfo {author} {\bibfnamefont {J.}~\bibnamefont {Balgley}}, \bibinfo {author} {\bibfnamefont {J.}~\bibnamefont {Butler}}, \bibinfo {author} {\bibfnamefont {S.}~\bibnamefont {Biswas}}, \bibinfo {author} {\bibfnamefont {Z.}~\bibnamefont {Ge}}, \bibinfo {author} {\bibfnamefont {S.}~\bibnamefont {Lagasse}}, \bibinfo {author} {\bibfnamefont {T.}~\bibnamefont {Taniguchi}}, \bibinfo {author} {\bibfnamefont {K.}~\bibnamefont {Watanabe}}, \bibinfo {author} {\bibfnamefont {M.}~\bibnamefont {Cothrine}}, \bibinfo {author} {\bibfnamefont {D.~G.}\ \bibnamefont {Mandrus}}, \bibinfo {author} {\bibfnamefont {J.}~\bibnamefont {Velasco~Jr}}, \emph {et~al.},\ }\bibfield  {title} {\bibinfo {title} {{Ultrasharp Lateral p--n Junctions in PModulation-Doped Graphene}},\ }\href@noop {} {\bibfield  {journal} {\bibinfo  {journal} {Nano. Lett.}\ }\textbf {\bibinfo {volume} {22}},\ \bibinfo {pages} {4124} (\bibinfo {year} {2022})}\BibitemShut {NoStop}%
\bibitem [{\citenamefont {Xu}\ and\ \citenamefont {Lai}(2016)}]{xu:2016}%
  \BibitemOpen
  \bibfield  {author} {\bibinfo {author} {\bibfnamefont {H.-Y.}\ \bibnamefont {Xu}}\ and\ \bibinfo {author} {\bibfnamefont {Y.-C.}\ \bibnamefont {Lai}},\ }\bibfield  {title} {\bibinfo {title} {Revival resonant scattering, perfect caustics, and isotropic transport of pseudospin-1 particles},\ }\href {https://doi.org/10.1103/PhysRevB.94.165405} {\bibfield  {journal} {\bibinfo  {journal} {Phys. Rev. B}\ }\textbf {\bibinfo {volume} {94}},\ \bibinfo {pages} {165405} (\bibinfo {year} {2016})}\BibitemShut {NoStop}%
\bibitem [{\citenamefont {\ifmmode \check{Z}\else \v{Z}\fi{}uti\ifmmode~\acute{c}\else \'{c}\fi{}}\ \emph {et~al.}(2004)\citenamefont {\ifmmode \check{Z}\else \v{Z}\fi{}uti\ifmmode~\acute{c}\else \'{c}\fi{}}, \citenamefont {Fabian},\ and\ \citenamefont {Das~Sarma}}]{vzutic:2004}%
  \BibitemOpen
  \bibfield  {author} {\bibinfo {author} {\bibfnamefont {I.}~\bibnamefont {\ifmmode \check{Z}\else \v{Z}\fi{}uti\ifmmode~\acute{c}\else \'{c}\fi{}}}, \bibinfo {author} {\bibfnamefont {J.}~\bibnamefont {Fabian}},\ and\ \bibinfo {author} {\bibfnamefont {S.}~\bibnamefont {Das~Sarma}},\ }\bibfield  {title} {\bibinfo {title} {Spintronics: Fundamentals and applications},\ }\href {https://doi.org/10.1103/RevModPhys.76.323} {\bibfield  {journal} {\bibinfo  {journal} {Rev. Mod. Phys.}\ }\textbf {\bibinfo {volume} {76}},\ \bibinfo {pages} {323} (\bibinfo {year} {2004})}\BibitemShut {NoStop}%
\bibitem [{\citenamefont {Dieny}\ \emph {et~al.}(2020)\citenamefont {Dieny}, \citenamefont {Prejbeanu}, \citenamefont {Garello}, \citenamefont {Gambardella}, \citenamefont {Freitas}, \citenamefont {Lehndorff}, \citenamefont {Raberg}, \citenamefont {Ebels}, \citenamefont {Demokritov}, \citenamefont {Akerman} \emph {et~al.}}]{dieny:2020}%
  \BibitemOpen
  \bibfield  {author} {\bibinfo {author} {\bibfnamefont {B.}~\bibnamefont {Dieny}}, \bibinfo {author} {\bibfnamefont {I.~L.}\ \bibnamefont {Prejbeanu}}, \bibinfo {author} {\bibfnamefont {K.}~\bibnamefont {Garello}}, \bibinfo {author} {\bibfnamefont {P.}~\bibnamefont {Gambardella}}, \bibinfo {author} {\bibfnamefont {P.}~\bibnamefont {Freitas}}, \bibinfo {author} {\bibfnamefont {R.}~\bibnamefont {Lehndorff}}, \bibinfo {author} {\bibfnamefont {W.}~\bibnamefont {Raberg}}, \bibinfo {author} {\bibfnamefont {U.}~\bibnamefont {Ebels}}, \bibinfo {author} {\bibfnamefont {S.~O.}\ \bibnamefont {Demokritov}}, \bibinfo {author} {\bibfnamefont {J.}~\bibnamefont {Akerman}}, \emph {et~al.},\ }\bibfield  {title} {\bibinfo {title} {Opportunities and challenges for spintronics in the microelectronics industry},\ }\href@noop {} {\bibfield  {journal} {\bibinfo  {journal} {Nat. Electron.}\ }\textbf {\bibinfo {volume} {3}},\ \bibinfo {pages} {446} (\bibinfo {year} {2020})}\BibitemShut {NoStop}%
\bibitem [{\citenamefont {Datta}\ and\ \citenamefont {Das}(1990)}]{datta:1990}%
  \BibitemOpen
  \bibfield  {author} {\bibinfo {author} {\bibfnamefont {S.}~\bibnamefont {Datta}}\ and\ \bibinfo {author} {\bibfnamefont {B.}~\bibnamefont {Das}},\ }\bibfield  {title} {\bibinfo {title} {Electronic analog of the electro-optic modulator},\ }\href@noop {} {\bibfield  {journal} {\bibinfo  {journal} {Appl. Phys. Lett.}\ }\textbf {\bibinfo {volume} {56}},\ \bibinfo {pages} {665} (\bibinfo {year} {1990})}\BibitemShut {NoStop}%
\bibitem [{\citenamefont {Chuang}\ \emph {et~al.}(2015)\citenamefont {Chuang}, \citenamefont {Ho}, \citenamefont {Smith}, \citenamefont {Sfigakis}, \citenamefont {Pepper}, \citenamefont {Chen}, \citenamefont {Fan}, \citenamefont {Griffiths}, \citenamefont {Farrer}, \citenamefont {Beere} \emph {et~al.}}]{chuang:2015}%
  \BibitemOpen
  \bibfield  {author} {\bibinfo {author} {\bibfnamefont {P.}~\bibnamefont {Chuang}}, \bibinfo {author} {\bibfnamefont {S.-C.}\ \bibnamefont {Ho}}, \bibinfo {author} {\bibfnamefont {L.~W.}\ \bibnamefont {Smith}}, \bibinfo {author} {\bibfnamefont {F.}~\bibnamefont {Sfigakis}}, \bibinfo {author} {\bibfnamefont {M.}~\bibnamefont {Pepper}}, \bibinfo {author} {\bibfnamefont {C.-H.}\ \bibnamefont {Chen}}, \bibinfo {author} {\bibfnamefont {J.-C.}\ \bibnamefont {Fan}}, \bibinfo {author} {\bibfnamefont {J.}~\bibnamefont {Griffiths}}, \bibinfo {author} {\bibfnamefont {I.}~\bibnamefont {Farrer}}, \bibinfo {author} {\bibfnamefont {H.~E.}\ \bibnamefont {Beere}}, \emph {et~al.},\ }\bibfield  {title} {\bibinfo {title} {All-electric all-semiconductor spin field-effect transistors},\ }\href@noop {} {\bibfield  {journal} {\bibinfo  {journal} {Nat. Nanotechnol.}\ }\textbf {\bibinfo {volume} {10}},\ \bibinfo {pages} {35} (\bibinfo {year} {2015})}\BibitemShut {NoStop}%
\bibitem [{\citenamefont {Yan}\ \emph {et~al.}(2016)\citenamefont {Yan}, \citenamefont {Txoperena}, \citenamefont {Llopis}, \citenamefont {Dery}, \citenamefont {Hueso},\ and\ \citenamefont {Casanova}}]{yan:2016}%
  \BibitemOpen
  \bibfield  {author} {\bibinfo {author} {\bibfnamefont {W.}~\bibnamefont {Yan}}, \bibinfo {author} {\bibfnamefont {O.}~\bibnamefont {Txoperena}}, \bibinfo {author} {\bibfnamefont {R.}~\bibnamefont {Llopis}}, \bibinfo {author} {\bibfnamefont {H.}~\bibnamefont {Dery}}, \bibinfo {author} {\bibfnamefont {L.~E.}\ \bibnamefont {Hueso}},\ and\ \bibinfo {author} {\bibfnamefont {F.}~\bibnamefont {Casanova}},\ }\bibfield  {title} {\bibinfo {title} {A two-dimensional spin field-effect switch},\ }\href@noop {} {\bibfield  {journal} {\bibinfo  {journal} {Nat. Commun.}\ }\textbf {\bibinfo {volume} {7}},\ \bibinfo {pages} {13372} (\bibinfo {year} {2016})}\BibitemShut {NoStop}%
\bibitem [{\citenamefont {Jiang}\ \emph {et~al.}(2019)\citenamefont {Jiang}, \citenamefont {Li}, \citenamefont {Wang}, \citenamefont {Shan},\ and\ \citenamefont {Mak}}]{jiang:2019}%
  \BibitemOpen
  \bibfield  {author} {\bibinfo {author} {\bibfnamefont {S.}~\bibnamefont {Jiang}}, \bibinfo {author} {\bibfnamefont {L.}~\bibnamefont {Li}}, \bibinfo {author} {\bibfnamefont {Z.}~\bibnamefont {Wang}}, \bibinfo {author} {\bibfnamefont {J.}~\bibnamefont {Shan}},\ and\ \bibinfo {author} {\bibfnamefont {K.~F.}\ \bibnamefont {Mak}},\ }\bibfield  {title} {\bibinfo {title} {Spin tunnel field-effect transistors based on two-dimensional van der waals heterostructures},\ }\href@noop {} {\bibfield  {journal} {\bibinfo  {journal} {Nat. Electron.}\ }\textbf {\bibinfo {volume} {2}},\ \bibinfo {pages} {159} (\bibinfo {year} {2019})}\BibitemShut {NoStop}%
\bibitem [{\citenamefont {Malik}\ \emph {et~al.}(2020)\citenamefont {Malik}, \citenamefont {Kharadi}, \citenamefont {Khanday},\ and\ \citenamefont {Parveen}}]{malik:2020}%
  \BibitemOpen
  \bibfield  {author} {\bibinfo {author} {\bibfnamefont {G.~F.~A.}\ \bibnamefont {Malik}}, \bibinfo {author} {\bibfnamefont {M.~A.}\ \bibnamefont {Kharadi}}, \bibinfo {author} {\bibfnamefont {F.~A.}\ \bibnamefont {Khanday}},\ and\ \bibinfo {author} {\bibfnamefont {N.}~\bibnamefont {Parveen}},\ }\bibfield  {title} {\bibinfo {title} {Spin field effect transistors and their applications: A survey},\ }\href@noop {} {\bibfield  {journal} {\bibinfo  {journal} {Microelectron. J.}\ }\textbf {\bibinfo {volume} {106}},\ \bibinfo {pages} {104924} (\bibinfo {year} {2020})}\BibitemShut {NoStop}%
\bibitem [{\citenamefont {Liu}\ \emph {et~al.}(2021)\citenamefont {Liu}, \citenamefont {Peng}, \citenamefont {Cai}, \citenamefont {Yue}, \citenamefont {Wei}, \citenamefont {Impundu}, \citenamefont {Liu}, \citenamefont {Jin}, \citenamefont {Yang}, \citenamefont {Chu} \emph {et~al.}}]{liu:2021}%
  \BibitemOpen
  \bibfield  {author} {\bibinfo {author} {\bibfnamefont {J.}~\bibnamefont {Liu}}, \bibinfo {author} {\bibfnamefont {Z.}~\bibnamefont {Peng}}, \bibinfo {author} {\bibfnamefont {J.}~\bibnamefont {Cai}}, \bibinfo {author} {\bibfnamefont {J.}~\bibnamefont {Yue}}, \bibinfo {author} {\bibfnamefont {H.}~\bibnamefont {Wei}}, \bibinfo {author} {\bibfnamefont {J.}~\bibnamefont {Impundu}}, \bibinfo {author} {\bibfnamefont {H.}~\bibnamefont {Liu}}, \bibinfo {author} {\bibfnamefont {J.}~\bibnamefont {Jin}}, \bibinfo {author} {\bibfnamefont {Z.}~\bibnamefont {Yang}}, \bibinfo {author} {\bibfnamefont {W.}~\bibnamefont {Chu}}, \emph {et~al.},\ }\bibfield  {title} {\bibinfo {title} {A room-temperature four-terminal spin field effect transistor},\ }\href@noop {} {\bibfield  {journal} {\bibinfo  {journal} {Nano Today}\ }\textbf {\bibinfo {volume} {38}},\ \bibinfo {pages} {101138} (\bibinfo {year} {2021})}\BibitemShut {NoStop}%
\bibitem [{\citenamefont {Han}\ \emph {et~al.}(2014)\citenamefont {Han}, \citenamefont {Kawakami}, \citenamefont {Gmitra},\ and\ \citenamefont {Fabian}}]{han:2014}%
  \BibitemOpen
  \bibfield  {author} {\bibinfo {author} {\bibfnamefont {W.}~\bibnamefont {Han}}, \bibinfo {author} {\bibfnamefont {R.~K.}\ \bibnamefont {Kawakami}}, \bibinfo {author} {\bibfnamefont {M.}~\bibnamefont {Gmitra}},\ and\ \bibinfo {author} {\bibfnamefont {J.}~\bibnamefont {Fabian}},\ }\bibfield  {title} {\bibinfo {title} {Graphene spintronics},\ }\href@noop {} {\bibfield  {journal} {\bibinfo  {journal} {Nat. Nanotechnol.}\ }\textbf {\bibinfo {volume} {9}},\ \bibinfo {pages} {794} (\bibinfo {year} {2014})}\BibitemShut {NoStop}%
\bibitem [{\citenamefont {Tombros}\ \emph {et~al.}(2007)\citenamefont {Tombros}, \citenamefont {Jozsa}, \citenamefont {Popinciuc}, \citenamefont {Jonkman},\ and\ \citenamefont {Van~Wees}}]{tombros:2007}%
  \BibitemOpen
  \bibfield  {author} {\bibinfo {author} {\bibfnamefont {N.}~\bibnamefont {Tombros}}, \bibinfo {author} {\bibfnamefont {C.}~\bibnamefont {Jozsa}}, \bibinfo {author} {\bibfnamefont {M.}~\bibnamefont {Popinciuc}}, \bibinfo {author} {\bibfnamefont {H.~T.}\ \bibnamefont {Jonkman}},\ and\ \bibinfo {author} {\bibfnamefont {B.~J.}\ \bibnamefont {Van~Wees}},\ }\bibfield  {title} {\bibinfo {title} {Electronic spin transport and spin precession in single graphene layers at room temperature},\ }\href@noop {} {\bibfield  {journal} {\bibinfo  {journal} {Nature}\ }\textbf {\bibinfo {volume} {448}},\ \bibinfo {pages} {571} (\bibinfo {year} {2007})}\BibitemShut {NoStop}%
\bibitem [{\citenamefont {Yang}\ \emph {et~al.}(2011)\citenamefont {Yang}, \citenamefont {Balakrishnan}, \citenamefont {Volmer}, \citenamefont {Avsar}, \citenamefont {Jaiswal}, \citenamefont {Samm}, \citenamefont {Ali}, \citenamefont {Pachoud}, \citenamefont {Zeng}, \citenamefont {Popinciuc} \emph {et~al.}}]{yang:2011}%
  \BibitemOpen
  \bibfield  {author} {\bibinfo {author} {\bibfnamefont {T.-Y.}\ \bibnamefont {Yang}}, \bibinfo {author} {\bibfnamefont {J.}~\bibnamefont {Balakrishnan}}, \bibinfo {author} {\bibfnamefont {F.}~\bibnamefont {Volmer}}, \bibinfo {author} {\bibfnamefont {A.}~\bibnamefont {Avsar}}, \bibinfo {author} {\bibfnamefont {M.}~\bibnamefont {Jaiswal}}, \bibinfo {author} {\bibfnamefont {J.}~\bibnamefont {Samm}}, \bibinfo {author} {\bibfnamefont {S.}~\bibnamefont {Ali}}, \bibinfo {author} {\bibfnamefont {A.}~\bibnamefont {Pachoud}}, \bibinfo {author} {\bibfnamefont {M.}~\bibnamefont {Zeng}}, \bibinfo {author} {\bibfnamefont {M.}~\bibnamefont {Popinciuc}}, \emph {et~al.},\ }\bibfield  {title} {\bibinfo {title} {Observation of long spin-relaxation times in bilayer graphene at room temperature},\ }\href@noop {} {\bibfield  {journal} {\bibinfo  {journal} {Phys. Rev. Lett.}\ }\textbf {\bibinfo {volume} {107}},\ \bibinfo {pages} {047206} (\bibinfo {year} {2011})}\BibitemShut {NoStop}%
\bibitem [{\citenamefont {Han}\ \emph {et~al.}(2010)\citenamefont {Han}, \citenamefont {Pi}, \citenamefont {McCreary}, \citenamefont {Li}, \citenamefont {Wong}, \citenamefont {Swartz},\ and\ \citenamefont {Kawakami}}]{han:2010}%
  \BibitemOpen
  \bibfield  {author} {\bibinfo {author} {\bibfnamefont {W.}~\bibnamefont {Han}}, \bibinfo {author} {\bibfnamefont {K.}~\bibnamefont {Pi}}, \bibinfo {author} {\bibfnamefont {K.~M.}\ \bibnamefont {McCreary}}, \bibinfo {author} {\bibfnamefont {Y.}~\bibnamefont {Li}}, \bibinfo {author} {\bibfnamefont {J.~J.}\ \bibnamefont {Wong}}, \bibinfo {author} {\bibfnamefont {A.}~\bibnamefont {Swartz}},\ and\ \bibinfo {author} {\bibfnamefont {R.}~\bibnamefont {Kawakami}},\ }\bibfield  {title} {\bibinfo {title} {Tunneling spin injection into single layer graphene},\ }\href@noop {} {\bibfield  {journal} {\bibinfo  {journal} {Phys. Rev. Lett.}\ }\textbf {\bibinfo {volume} {105}},\ \bibinfo {pages} {167202} (\bibinfo {year} {2010})}\BibitemShut {NoStop}%
\bibitem [{\citenamefont {Dlubak}\ \emph {et~al.}(2012)\citenamefont {Dlubak}, \citenamefont {Martin}, \citenamefont {Deranlot}, \citenamefont {Servet}, \citenamefont {Xavier}, \citenamefont {Mattana}, \citenamefont {Sprinkle}, \citenamefont {Berger}, \citenamefont {De~Heer}, \citenamefont {Petroff} \emph {et~al.}}]{dlubak:2012}%
  \BibitemOpen
  \bibfield  {author} {\bibinfo {author} {\bibfnamefont {B.}~\bibnamefont {Dlubak}}, \bibinfo {author} {\bibfnamefont {M.-B.}\ \bibnamefont {Martin}}, \bibinfo {author} {\bibfnamefont {C.}~\bibnamefont {Deranlot}}, \bibinfo {author} {\bibfnamefont {B.}~\bibnamefont {Servet}}, \bibinfo {author} {\bibfnamefont {S.}~\bibnamefont {Xavier}}, \bibinfo {author} {\bibfnamefont {R.}~\bibnamefont {Mattana}}, \bibinfo {author} {\bibfnamefont {M.}~\bibnamefont {Sprinkle}}, \bibinfo {author} {\bibfnamefont {C.}~\bibnamefont {Berger}}, \bibinfo {author} {\bibfnamefont {W.~A.}\ \bibnamefont {De~Heer}}, \bibinfo {author} {\bibfnamefont {F.}~\bibnamefont {Petroff}}, \emph {et~al.},\ }\bibfield  {title} {\bibinfo {title} {Highly efficient spin transport in epitaxial graphene on {SiC}},\ }\href@noop {} {\bibfield  {journal} {\bibinfo  {journal} {Nat. Phys.}\ }\textbf {\bibinfo {volume} {8}},\ \bibinfo {pages} {557} (\bibinfo {year} {2012})}\BibitemShut {NoStop}%
\bibitem [{\citenamefont {Maksym}\ and\ \citenamefont {Aoki}(2021)}]{maksym:2021}%
  \BibitemOpen
  \bibfield  {author} {\bibinfo {author} {\bibfnamefont {P.}~\bibnamefont {Maksym}}\ and\ \bibinfo {author} {\bibfnamefont {H.}~\bibnamefont {Aoki}},\ }\bibfield  {title} {\bibinfo {title} {Complete spin and valley polarization by total external reflection from potential barriers in bilayer graphene and monolayer transition metal dichalcogenides},\ }\href@noop {} {\bibfield  {journal} {\bibinfo  {journal} {Physical Review B}\ }\textbf {\bibinfo {volume} {104}},\ \bibinfo {pages} {155401} (\bibinfo {year} {2021})}\BibitemShut {NoStop}%
\bibitem [{\citenamefont {Zeng}\ \emph {et~al.}(2011)\citenamefont {Zeng}, \citenamefont {Shen}, \citenamefont {Zhou}, \citenamefont {Zhang}, \citenamefont {Feng} \emph {et~al.}}]{zeng:2011}%
  \BibitemOpen
  \bibfield  {author} {\bibinfo {author} {\bibfnamefont {M.}~\bibnamefont {Zeng}}, \bibinfo {author} {\bibfnamefont {L.}~\bibnamefont {Shen}}, \bibinfo {author} {\bibfnamefont {M.}~\bibnamefont {Zhou}}, \bibinfo {author} {\bibfnamefont {C.}~\bibnamefont {Zhang}}, \bibinfo {author} {\bibfnamefont {Y.}~\bibnamefont {Feng}}, \emph {et~al.},\ }\bibfield  {title} {\bibinfo {title} {Graphene-based bipolar spin diode and spin transistor: Rectification and amplification of spin-polarized current},\ }\href@noop {} {\bibfield  {journal} {\bibinfo  {journal} {Phys. Rev. B}\ }\textbf {\bibinfo {volume} {83}},\ \bibinfo {pages} {115427} (\bibinfo {year} {2011})}\BibitemShut {NoStop}%
\bibitem [{\citenamefont {Dugaev}\ \emph {et~al.}(2006)\citenamefont {Dugaev}, \citenamefont {Litvinov},\ and\ \citenamefont {Barnas}}]{dugaev:2006}%
  \BibitemOpen
  \bibfield  {author} {\bibinfo {author} {\bibfnamefont {V.~K.}\ \bibnamefont {Dugaev}}, \bibinfo {author} {\bibfnamefont {V.~I.}\ \bibnamefont {Litvinov}},\ and\ \bibinfo {author} {\bibfnamefont {J.}~\bibnamefont {Barnas}},\ }\bibfield  {title} {\bibinfo {title} {Exchange interaction of magnetic impurities in graphene},\ }\href {https://doi.org/10.1103/PhysRevB.74.224438} {\bibfield  {journal} {\bibinfo  {journal} {Phys. Rev. B}\ }\textbf {\bibinfo {volume} {74}},\ \bibinfo {pages} {224438} (\bibinfo {year} {2006})}\BibitemShut {NoStop}%
\bibitem [{\citenamefont {Jiang}\ \emph {et~al.}(2017)\citenamefont {Jiang}, \citenamefont {Mao}, \citenamefont {Moldovan}, \citenamefont {Masir}, \citenamefont {Li}, \citenamefont {Watanabe}, \citenamefont {Taniguchi}, \citenamefont {Peeters},\ and\ \citenamefont {Andrei}}]{jiang:2017}%
  \BibitemOpen
  \bibfield  {author} {\bibinfo {author} {\bibfnamefont {Y.}~\bibnamefont {Jiang}}, \bibinfo {author} {\bibfnamefont {J.}~\bibnamefont {Mao}}, \bibinfo {author} {\bibfnamefont {D.}~\bibnamefont {Moldovan}}, \bibinfo {author} {\bibfnamefont {M.~R.}\ \bibnamefont {Masir}}, \bibinfo {author} {\bibfnamefont {G.}~\bibnamefont {Li}}, \bibinfo {author} {\bibfnamefont {K.}~\bibnamefont {Watanabe}}, \bibinfo {author} {\bibfnamefont {T.}~\bibnamefont {Taniguchi}}, \bibinfo {author} {\bibfnamefont {F.~M.}\ \bibnamefont {Peeters}},\ and\ \bibinfo {author} {\bibfnamefont {E.~Y.}\ \bibnamefont {Andrei}},\ }\bibfield  {title} {\bibinfo {title} {Tuning a circular p--n junction in graphene from quantum confinement to optical guiding},\ }\href@noop {} {\bibfield  {journal} {\bibinfo  {journal} {Nat. Nanotechnol.}\ }\textbf {\bibinfo {volume} {12}},\ \bibinfo {pages} {1045} (\bibinfo {year} {2017})}\BibitemShut {NoStop}%
\bibitem [{\citenamefont {Wang}\ and\ \citenamefont {Ran}(2011)}]{Wang:2011}%
  \BibitemOpen
  \bibfield  {author} {\bibinfo {author} {\bibfnamefont {F.}~\bibnamefont {Wang}}\ and\ \bibinfo {author} {\bibfnamefont {Y.}~\bibnamefont {Ran}},\ }\bibfield  {title} {\bibinfo {title} {Nearly flat band with chern number $c=2$ on the dice lattice},\ }\href {https://doi.org/10.1103/PhysRevB.84.241103} {\bibfield  {journal} {\bibinfo  {journal} {Phys. Rev. B}\ }\textbf {\bibinfo {volume} {84}},\ \bibinfo {pages} {241103} (\bibinfo {year} {2011})}\BibitemShut {NoStop}%
\bibitem [{\citenamefont {Romh{\'a}nyi}\ \emph {et~al.}(2015)\citenamefont {Romh{\'a}nyi}, \citenamefont {Penc},\ and\ \citenamefont {Ganesh}}]{romhanyi:2015}%
  \BibitemOpen
  \bibfield  {author} {\bibinfo {author} {\bibfnamefont {J.}~\bibnamefont {Romh{\'a}nyi}}, \bibinfo {author} {\bibfnamefont {K.}~\bibnamefont {Penc}},\ and\ \bibinfo {author} {\bibfnamefont {R.}~\bibnamefont {Ganesh}},\ }\bibfield  {title} {\bibinfo {title} {Hall effect of triplons in a dimerized quantum magnet},\ }\href@noop {} {\bibfield  {journal} {\bibinfo  {journal} {Nat. Commun.}\ }\textbf {\bibinfo {volume} {6}},\ \bibinfo {pages} {6805} (\bibinfo {year} {2015})}\BibitemShut {NoStop}%
\bibitem [{\citenamefont {Giovannetti}\ \emph {et~al.}(2015)\citenamefont {Giovannetti}, \citenamefont {Capone}, \citenamefont {van~den Brink},\ and\ \citenamefont {Ortix}}]{giovannetti:2015}%
  \BibitemOpen
  \bibfield  {author} {\bibinfo {author} {\bibfnamefont {G.}~\bibnamefont {Giovannetti}}, \bibinfo {author} {\bibfnamefont {M.}~\bibnamefont {Capone}}, \bibinfo {author} {\bibfnamefont {J.}~\bibnamefont {van~den Brink}},\ and\ \bibinfo {author} {\bibfnamefont {C.}~\bibnamefont {Ortix}},\ }\bibfield  {title} {\bibinfo {title} {Kekul{\'e} textures, pseudospin-one {Dirac} cones, and quadratic band crossings in a graphene-hexagonal indium chalcogenide bilayer},\ }\href@noop {} {\bibfield  {journal} {\bibinfo  {journal} {Phys. Rev. B}\ }\textbf {\bibinfo {volume} {91}},\ \bibinfo {pages} {121417} (\bibinfo {year} {2015})}\BibitemShut {NoStop}%
\bibitem [{\citenamefont {Tan}\ and\ \citenamefont {Jalil}(2012)}]{TAN2012141}%
  \BibitemOpen
  \bibfield  {author} {\bibinfo {author} {\bibfnamefont {S.~G.}\ \bibnamefont {Tan}}\ and\ \bibinfo {author} {\bibfnamefont {M.~B.}\ \bibnamefont {Jalil}},\ }\bibfield  {title} {\bibinfo {title} {5 - spintronics and spin {Hall} effects in nanoelectronics},\ }in\ \href {https://doi.org/https://doi.org/10.1533/9780857095886.141} {\emph {\bibinfo {booktitle} {Introduction to the Physics of Nanoelectronics}}},\ \bibinfo {series and number} {Woodhead Publishing Series in Electronic and Optical Materials},\ \bibinfo {editor} {edited by\ \bibinfo {editor} {\bibfnamefont {S.~G.}\ \bibnamefont {Tan}}\ and\ \bibinfo {editor} {\bibfnamefont {M.~B.}\ \bibnamefont {Jalil}}}\ (\bibinfo  {publisher} {Woodhead Publishing},\ \bibinfo {year} {2012})\ pp.\ \bibinfo {pages} {141--197}\BibitemShut {NoStop}%
\bibitem [{\citenamefont {Curtright}\ \emph {et~al.}(2014)\citenamefont {Curtright}, \citenamefont {Fairlie}, \citenamefont {Zachos} \emph {et~al.}}]{curtright:2014}%
  \BibitemOpen
  \bibfield  {author} {\bibinfo {author} {\bibfnamefont {T.~L.}\ \bibnamefont {Curtright}}, \bibinfo {author} {\bibfnamefont {D.~B.}\ \bibnamefont {Fairlie}}, \bibinfo {author} {\bibfnamefont {C.~K.}\ \bibnamefont {Zachos}}, \emph {et~al.},\ }\bibfield  {title} {\bibinfo {title} {A compact formula for rotations as spin matrix polynomials},\ }\href@noop {} {\bibfield  {journal} {\bibinfo  {journal} {SIGMA. Symmetry, Integrability and Geometry: Methods and Applications}\ }\textbf {\bibinfo {volume} {10}},\ \bibinfo {pages} {084} (\bibinfo {year} {2014})}\BibitemShut {NoStop}%
\bibitem [{\citenamefont {{Wikipedia contributors}}(2022)}]{enwiki:1116845797}%
  \BibitemOpen
  \bibfield  {author} {\bibinfo {author} {\bibnamefont {{Wikipedia contributors}}},\ }\href {https://en.wikipedia.org/w/index.php?title=Momentum-transfer_cross_section&oldid=1116845797} {\bibinfo {title} {Momentum-transfer cross section --- {Wikipedia}{,} {The Free Encyclopedia}}} (\bibinfo {year} {2022}),\ \bibinfo {note} {[Online; accessed 30-January-2023]}\BibitemShut {NoStop}%
\bibitem [{\citenamefont {Wiersig}\ and\ \citenamefont {Hentschel}(2008)}]{wiersig:2008}%
  \BibitemOpen
  \bibfield  {author} {\bibinfo {author} {\bibfnamefont {J.}~\bibnamefont {Wiersig}}\ and\ \bibinfo {author} {\bibfnamefont {M.}~\bibnamefont {Hentschel}},\ }\bibfield  {title} {\bibinfo {title} {Combining directional light output and ultralow loss in deformed microdisks},\ }\href {https://doi.org/10.1103/PhysRevLett.100.033901} {\bibfield  {journal} {\bibinfo  {journal} {Phys. Rev. Lett.}\ }\textbf {\bibinfo {volume} {100}},\ \bibinfo {pages} {033901} (\bibinfo {year} {2008})}\BibitemShut {NoStop}%
\end{thebibliography}%

\end{document}